\documentclass[useAMS,usenatbib]{mn2e}
\pdfoutput=1
\usepackage{amssymb}
\usepackage{aas_macros}
\usepackage{natbib}
\usepackage{times}
\usepackage{graphicx}
\newcommand{\gimic}          {\textsc{Gimic}}
\newcommand{\gadget}         {\textsc{Gadget3}}
\newcommand{\owls}         {\textsc{Owls}}

\renewcommand{\sp}           {\hbox{$+1\sigma$}}
\newcommand{\spp}            {\hbox{$+2\sigma$}}
\newcommand{\sm}             {\hbox{$-1\sigma$}}
\newcommand{\smm}            {\hbox{$-2\sigma$}}
                             
\newcommand{\yr}             {\,{\rm yr}}

\newcommand{\Gyr}            {\,{\rm Gyr}}
\newcommand{\perGyr}         {\,{\rm Gyr}^{-1}}
                             
\newcommand{\pc}             {\,{\rm pc}}
\newcommand{\kpc}            {\,{\rm kpc}}

\newcommand{\Msun}           {\,{\rm M}_\odot}
\newcommand{\Msunyr}         {\,{\rm M}_\odot\,{\rm yr}^{-1}}
\newcommand{\MsunyrMpccubed} {\,{\rm M}_\odot\,{\rm yr}^{-1}\,{\rm Mpc}^{-3}}
\newcommand{\hkpc}           {\,h^{-1}\,{\rm kpc}}
\newcommand{\hMpc}           {\,h^{-1}\,{\rm Mpc}}
\newcommand{\hMsun}          {\,h^{-1}\,{\rm M}_\odot}
\newcommand{\kms}            {\,\,{\rm km}\,\,{\rm s}^{-1}}
\newcommand{\HubUnits}     {\,\,{\rm km}\,\,{\rm s}^{-1}\,{\rm Mpc}^{-1}}
\newcommand{\cmcubed}        {\,\,{\rm cm}^{-3}}

\newcommand{\Zsun}           {\,{\rm Z}_\odot}
\newcommand{\erg}            {\,{\rm erg}}
\newcommand{\K}              {\,{\rm K}}
\newcommand{\sfrd}          {$\dot{\rho}_\star$}
\newcommand{\sfrdz}          {$\dot{\rho}_\star(z)$}

\def\lsim{ \lower .75ex \hbox{$\sim$} \llap{\raise .27ex \hbox{$<$}}}
\def\gsim{ \lower .75ex \hbox{$\sim$} \llap{\raise .27ex \hbox{$>$}}}

\title[GIMIC]{Galaxies-Intergalactic Medium Interaction Calculation
  \\--I. Galaxy formation as a function of large-scale environment}

\author[R.~A.~Crain et al.]  {\parbox[h]{160mm} { Robert A.
    Crain$^{1,2}$\thanks{E-mail: rcrain@astro.swin.edu.au}, Tom
    Theuns$^{1,3}$, Claudio Dalla Vecchia$^4$, Vincent R. Eke$^1$,
    Carlos S. Frenk$^1$, Adrian Jenkins$^1$, Scott T. Kay$^5$, John A.
    Peacock$^6$ Frazer R. Pearce$^7$, Joop Schaye$^4$, Volker
    Springel$^8$, Peter A. Thomas$^9$, Simon D. M. White$^8$ \& Robert
    P. C. Wiersma$^4$ (The Virgo Consortium)}
  \vspace{6pt}\\
  $^1$Institute for Computational Cosmology, Department of Physics, University of Durham, South Road, Durham, DH1 3LE, UK\\
  $^2$Centre for Astrophysics \& Supercomputing, Mail No. H39, Swinburne University of Technology, PO Box 218, Hawthorn, Victoria 3122, Australia\\
  $^3$Department of Physics, University of Antwerp, Campus Groenenborger, Groenenborgerlaan 171, B-2020 Antwerp, Belgium\\
  $^4$Leiden Observatory, Leiden University, PO Box 9513, 2300 RA Leiden, Netherlands\\
  $^5$Jodrell Bank Centre for Astrophysics, School of Physics \& Astronomy, Alan Turing Building, University of Manchester, Manchester, M13 9PL, UK\\
  $^6$Institute for Astronomy, University of Edinburgh, Royal Observatory, Edinburgh, EH9 3HJ, UK\\
  $^7$Department of Physics and Astronomy, University of Nottingham, Nottingham, NG7 2RD, UK\\
  $^8$Max-Planck-Institut f\"ur Astrophysik, Karl-Schwarzchild-Stra\ss{}e 1, 85740 Garching bei M\"{u}nchen, Germany\\
  $^{9}$Astronomy Centre, University of Sussex, Falmer, Brighton, BN1 9QH, UK}

\begin{document}

\date{\today}
\pagerange{\pageref{firstpage}--\pageref{lastpage}} \pubyear{2009}

\maketitle

\label{firstpage}

\begin{abstract}
We present the first results of hydrodynamical simulations that follow
the formation of galaxies to the present day in nearly spherical
regions of radius $\sim 20\hMpc$ drawn from the Millennium Simulation
(Springel et al.). The regions have mean overdensities that deviate by
$(-2$, $-1$, $0$, $+1$,$+2)\sigma$ from the cosmic mean, where
$\sigma$ is the {\em rms} mass fluctuation on a scale of $\sim20\hMpc$
at $z=1.5$.  The simulations have mass resolution of up to $\sim
10^6\hMsun$, cover the entire range of large-scale
cosmological environments, including rare objects such as massive
clusters and sparse voids, and allow extrapolation of statistics to
the $(500\hMpc)^3$ Millennium Simulation volume as a
whole. They include gas cooling, photoheating from an imposed ionising
background, supernova feedback and galactic winds, but no AGN.  In
this paper we focus on the star formation properties of the model.  We
find that the specific star formation rate density at $z\lesssim 10$
varies systematically from region to region by up to an order of
magnitude, but the global value, averaged over all volumes, closely
reproduces observational data. Massive, compact galaxies, similar to
those observed in the GOODS fields (Wiklind et al.), form in the
overdense regions as early as $z=6$, but do not appear in the
underdense regions until $z\sim 3$. These environmental variations are
not caused by a dependence of the star formation properties on
environment, but rather by a strong variation of the halo mass
function from one environment to another, with more massive haloes
forming preferentially in the denser regions. At all epochs, stars
form most efficiently in haloes of circular velocity $v_{\rm c} \sim
250\kms$. However, the star-formation history exhibits a
form of ``downsizing'' (even in the absence of AGN feedback): the stars comprising massive galaxies at $z=0$ have mostly formed by $z=1-2$, whilst those comprising smaller galaxies typically form at later times. However, additional
feedback is required to limit star formation in massive galaxies at late times.
\end{abstract}

\section{Introduction}

Numerical simulations have emerged, over the past two decades or so,
as a useful technique for modelling the formation and evolution of
cosmic structures. In particular,
they have yielded accurate predictions for the clustering and
evolution of dark matter, a relatively simple problem in which the
dynamics are determined purely by gravitational forces
\citep[e.g.][]{DEFW85}. Modelling the evolution of small-scale
structures involving baryons is considerably more challenging because
of the complexity of the physics involved and the greater
uncertainties in the modelling techniques
\citep[e.g.][]{Katz_and_Gunn_91}. In the regime where baryonic
processes are non-negligible, the ability of numerical simulations to
yield unique and robust predictions is therefore reduced. This is
particularly true for systems in which radiative cooling is efficient.

The inherent difficulties and computational expense of modelling
baryons in this regime has stimulated the development of semi-analytic
techniques
\citep{White_and_Frenk_91,Kauffmann_White_and_Guiderdoni_93,Cole_et_al_94,Cole_et_al_00,
  Somerville_and_Primack_99,Baugh_06}. These employ simplified
prescriptions for the evolution of baryons in dark matter haloes. The
formation histories of the haloes are followed using merger trees
constructed analytically with Monte Carlo techniques, or extracted
directly from $N$-body simulations. Notable examples of the latter
procedure are the studies of \citet{Bower_et_al_06},
\citet{Croton_et_al_06_short}, \citet{De_Lucia_et_al_06} and
\citet{Font_et_al_08_short}, based on merger histories derived from
the Millennium Simulation \citep{Springel_et_al_05_short}. The scale and
resolution of this dark matter simulation permits semi-analytical calculations
of the evolution of galaxies more massive than the Small Magellanic Cloud within
a comoving volume similar to that probed by the 2dFGRS
\citep{Colless_et_al_01_short} and SDSS
\citep{York_et_al_00_Short} at their median redshifts.

Simulations employing semi-analytic models have yielded critical
insights into the effects of baryonic processes on the observable
Universe. Nevertheless, the approximations inherent in this technique
limit the kind of processes that can be followed reliably in a
relatively simple manner. For example, while semi-analytic techniques
can be extended to study the evolution of the intergalactic medium
\citep[IGM; e.g.][]{Benson_et_al_02}, they are not well suited for studying
the interaction of gas with galactic ejecta, such as winds, and the
associated dynamical evolution. Processes of this kind can only be
studied in detail with simulations that explicitly follow the full
hydrodynamic evolution of the baryons \citep{Wiersma_et_al_09}.

The introduction and subsequent development of hydrodynamic numerical
methods within the framework of cosmological simulations has led to
important advances in our understanding of structure formation. For
example, the relationship between the Lyman-$\alpha$ forest and the
cosmic large-scale structure was first convincingly demonstrated using
such simulations
\citep[e.g.][]{Cen_et_al_94,Zhang_Anninos_and_Norman_95,Hernquist_et_al_96,Miralda-Escude_Cen_et_al_96,Zhang_et_al_97,Theuns_et_al_98,Dave_et_al_99}.
At the opposite end of the density scale, hydrodynamic simulations of
the formation of individual galaxies have confirmed the generic
requirement in hierarchical models (first identified using
semi-analytic techniques) for energy feedback mechanisms to prevent
the over-cooling of gas in small haloes at early times
\citep{White_and_Rees_78,Cole_91,White_and_Frenk_91}, and the associated
transfer of angular momentum from baryons to dark matter
\citep[e.g.][]{Navarro_and_Benz_91,Weil_Eke_and_Efstathiou_98,Thacker_and_Couchman_01,Abadi_Navarro_and_Steinmetz_03a,Abadi_Navarro_and_Steinmetz_03b,Sommer-Larsen_et_al_03,Governato_et_al_04_short,Robertson_et_al_04,Okamoto_et_al_05,Governato_et_al_07,Scannapieco_et_al_08,Zavala_Okamoto_and_Frenk_08}.

The traditional approach of hydrodynamical simulations has involved a
trade-off between resolution and computational volume. Such
trade-offs, however, are not possible if an unbiased statistical
sample of well resolved simulated galaxies is desired. This presents a
considerable computational challenge, and consequently previous
simulations have been limited to tracing volumes that are relatively
small in comparison with those used in large-scale structure
calculations and those probed by observational surveys. The gulf in
mass and spatial scale between individual galaxies and their
large-scale environment compounds the more fundamental dynamic range
problem faced by simulations of galaxy formation, that stems from the
influence of individual stars (via stellar winds and supernovae) on
gas.

Recent attempts to produce such samples include the smoothed particle
hydrodynamics (SPH) simulations of
\citet{Pearce_et_al_99_short,Pearce_et_al_01} using the \textsc{Hydra}
code \citep{Couchman_Thomas_and_Pearce_95}; those of
\citet{Dave_Finlator_Oppenheimer_06,Oppenheimer_and_Dave_06,Dave_and_Oppenheimer_07,Oppenheimer_and_Dave_07,Croft_et_al_08,Oppenheimer_and_Dave_08a},
using variants of the \textsc{Gadget2} code \citep{Springel_05}; and
those of \citet{Brooks_et_al_07} using the \textsc{Gasoline} code
\citep{Wadsley_Stadel_and_Quinn_04}. The adaptive mesh refinement (AMR)
technique for hydrodynamics has also been employed for simulations of
this type, for example by \citet[][see references therein for code
details]{Nagamine_et_al_05}, and by
\citet{Ocvirk_Pichon_and_Teyssier_08}, using the \textsc{Ramses} code
\citep{Teyssier_02}. To highlight an example, \citet{Croft_et_al_08} simulated (with $486^3$
particles of gas and dark matter) a periodic volume in which haloes of
Milky Way type galaxies were resolved with $\sim10^5$
particles. Although $\sim1000$ galaxies formed in this simulation, its
relatively small size (a cube of side $L=33.75\hMpc$) yielded few
massive galaxies and sampled a relatively narrow range of
cosmological environments. A further technical limitation of
relatively small volumes is that fluctuations on scales comparable to
the size of the box become non-negligible at a relatively high
redshift
\citep[][]{Bagla_and_Ray_05,Sirko_05}, after which the simulation
can no longer be considered a faithful representation of the underlying
model. For this reason, \citet{Croft_et_al_08} halted their simulation at $z=1$.

The need for detailed predictions of the properties of the low-redshift IGM and
the evolution of galaxies over the interval $0<z\lesssim1$ is highlighted by a
number of observational developments. The fate of the baryons seen in the
Lyman-$\alpha$ forest at high redshift is still largely uncharted but
measurements of the Lyman-$\alpha$ forest are beginning to probe the diffuse
baryons associated with the filaments of the cosmic web
\citep[e.g.][]{Pichon_et_al_01,Caucci_et_al_08}, where a significant fraction
may be in the form of a ``warm-hot intergalactic medium'' (WHIM) phase
\citep[e.g.][]{Cen_and_Ostriker_99a,Dave_et_al_99}. The recent
installation of the \textit{Cosmic Origins Spectrograph}
\citep{Green_Wilkinson_and_Morse_03} aboard the \textit{Hubble Space
Telescope} (HST), may yield significant advances in the understanding
of the low-redshift IGM. Alongside studies of the evolution of cosmic
gas, realistic modelling of the low-redshift evolution of galaxies is
essential to connect the snapshots of cosmic evolution provided by
galaxy surveys at different epochs, for example DEEP2
\citep{Davis_et_al_03_short}, VVDS \citep{LeFevre_et_al_05_short} and
COMBO-17 \citep{Wolf_et_al_03} at $z\sim 1$ with the 2dFGRS
and SDSS at $z\sim 0$.

Simulating large cosmological volumes ($L \gtrsim 100\hMpc$) at high resolution
($m_{\rm gas} \lesssim 10^7\hMsun$) to $z=0$ is infeasible with current
computational resources. To circumvent this fundamental limitation, we have
devised a new approach consisting of simulating at high resolution regions
extracted from the dark matter Millennium Simulation whose overdensities at
$z=1.5$ represent $(-2$, $-1$, $0$, $+1$,$+2)\sigma$ deviations from the cosmic
mean, where $\sigma$ is the rms mass fluctuation on a scale of
$\sim20\hMpc$. These simulations, the \textit{Galaxies-Intergalactic Medium
Interaction Calculation} (\textsc{Gimic}), follow the evolution of five roughly
spherical regions of radius $\sim20\hMpc$ from ``zoomed'' initial conditions
\citep{Frenk_et_al_96,Power_et_al_03,Navarro_et_al_04_short} using 
full gas dynamics, while the rest of the Millennium volume is resimulated at
lower resolution following dark matter only. We thus follow a very wide range of
cosmological environments whilst only simulating $0.13$ per cent of the
Millennium volume at high resolution. The extent of the dynamic range allowed by
our procedure is illustrated in Fig.~\ref{fig:VisStages} which, starting from
the full Millennium Simulation volume ($L=500\hMpc$), zooms in, first by a
factor of 10 to show a full \textsc{Gimic} sphere in the central panel, and then
by a further factor of 1000 to show a disc galaxy within this region.

In contrast to simulations of small periodic volumes, our adopted
strategy allows us to follow the cosmic evolution to $z=0$, since
fluctuations on the scale of the Millennium Simulation are still well
described by linear theory. Judicious selection of the five
\textsc{Gimic} regions across environmentally diverse regions yields a
sample spanning the range of structures found within the Millennium
Simulation. In our highest resolution realisations, the Jeans scale in
the IGM after the epoch of reionisation is marginally resolved. A further
advantage of resimulating regions of the Millennium Simulation is that
they complement the existing semi-analytic calculations implemented on
it
\citep{Bower_et_al_06,Croton_et_al_06_short,De_Lucia_et_al_06,Font_et_al_08_short}.

The simulations were performed using \textsc{Gadget3}, a significantly
upgraded variant of the \textsc{Gadget2} code described by
\cite{Springel_05}, and including new treatments of baryonic
processes, such as radiative cooling, heating by the metagalactic
ultraviolet background radiation, star formation, stellar feedback and
chemodynamics. The simulations are computationally expensive, and so
we have limited ourselves to running them with a single implementation
of the code. Analyses of these simulations will therefore, in general,
focus on the environmental effects that these simulations are able to
explore; we do not examine the role of, for example, changes to the
assumed initial mass function (IMF) or the implementation of
feedback. Such tests will be explored in the closely related
Overwhelmingly Large Simulations (\owls; Schaye et al.,
\textit{in prep}) project. In this, the first paper in a series, we focus on one
central aspect of these simulations, the difference in the star formation rate
density amongst the five different environments, and on how these 
combine to produce a ``cosmic'' star formation rate density
\citep[e.g.][]{Lilly_et_al_96,Madau_et_al_96}.

The paper is laid out as follows. In Section \ref{sec:methodology} we
describe the simulations, concentrating on new aspects of our
code, and provide a brief overview of the generation of the initial
conditions.  Our suite of simulations includes runs with varying
resolution to enable us to check for numerical convergence. In Section
\ref{sec:haloes}, we discuss the halo mass functions of the five
\textsc{Gimic} regions, which are very different from each other. In
Section \ref{sec:sfrd}, we calculate the evolution of the star
formation rate of each region and present an estimate for the entire
Millennium Simulation based on a weighted average of these
measurements. The star formation properties of haloes and their
galaxies are discussed in Section \ref{sec:halo_sfr}. In Section 6, we
carry out a limited comparison with observational data. We summarise
and conclude in Section~7. 

The simulations adopt the same cosmological parameters as the
Millennium Simulation: $\Omega_{\rm m} = 0.25$, $\Omega_\Lambda =
0.75$, $\Omega_{\rm b} = 0.045$, $n_{\rm s}=1$, $\sigma_8 = 0.9$, $H_0
= 100~h~{\rm km~s}^{-1}~{\rm Mpc}^{-1}$, $h=0.73$. This work is part
of the programme of the Virgo consortium for cosmological simulations.

\section{Methodology}
\label{sec:methodology}

This section describes the three central aspects of our methodology: the code
used to perform the simulations, the setup of the initial conditions, and the
identification of dark matter haloes and galaxies. Technical details of the
generation of the initial conditions are deferred to an Appendix. 
 
\subsection{Simulation code}
\label{sec:code}

For this study, we use \textsc{Gadget3}, an updated
version of the \textsc{Gadget2} code \citep{Springel_05}, to which
several physics modules have been added. The domain decomposition
differs from that in \textsc{Gadget2} in that it improves
load-balancing particularly for simulations (such as those described
here) with strongly clustered particle distributions run on parallel
supercomputers with a large number of cores. The hydrodynamics
implementation, taken from \textsc{Gadget2}, is the entropy conserving
formulation of SPH \citep{Gingold_and_Monaghan_77,Lucy_77}, as
discussed in \citet{Springel_and_Hernquist_2002}. SPH represents a
fluid by a set of particles that carry along a number of
internal properties (for example, mass, entropy), while other
properties are computed by interpolation over neighbouring particles
(for example, the density or pressure gradient). 

We have used new physics modules for star
formation \citep{Schaye_and_Dalla_Vecchia_08}, stellar feedback
\citep{Dalla_Vecchia_and_Schaye_08}, 
radiative cooling \citep{Wiersma_Schaye_and_Smith_09}, and
chemodynamics \citep{Wiersma_et_al_09}. Here we only provide
a brief overview of the main features of these modules.

\begin{itemize}
\item {\em Gas cooling and photoionisation.} Radiative cooling was
  implemented as described in
  \citet{Wiersma_Schaye_and_Smith_09}\footnote{We used their equation
  (3), rather than (4), and \textsc{cloudy} version 05.07.}. In brief,
  we assume the IGM is ionised and heated by a pervading uniform, but
  redshift dependent, ionising background from galaxies and quasars,
  as computed by \citet{Haardt_and_Madau_01}. We assume hydrogen
  reionises at redshift $z=9$, and helium II at $z = 3.5$
  \citep{Schaye_et_al_00, Theuns_et_al_02c}. The cooling rate is
  computed as a function of redshift, gas density, temperature and
  composition on an element-by-element basis using interpolation
  tables computed with \textsc{cloudy} \citep[as described
  in][]{Ferland_et_al_98}, assuming the gas to be optically thin and
  in ionisation equilibrium. During the reionisation of helium II, we
  inject 2~eV per proton, smoothed with a Gaussian function, centred
  at $z=3.5$ and with width $\sigma(z) = 0.5$, to mimic
  non-equilibrium and radiative-transfer effects
  \citep[e.g.][]{Abel_and_Haehnelt_99}. As demonstrated in \citet{Wiersma_et_al_09}, with this extra heat input the
  predicted temperature at the mean density is in good agreement with
  the measurements of \citet{Schaye_et_al_00}.

\item {\em Quiescent star formation and feedback.} 
 Cosmological simulations cannot, at present, resolve the multiphase structure
 of star-forming gas that arises from the combination of gas cooling and heating
 due to massive stars and supernovae (SNe). These processes make star-forming
 gas resistant to compression, since the energy output from SNe rapidly heats
 the interstellar medium (ISM), increasing its pressure. We mimic this by
 \textit{imposing} an effective equation of state $P=\kappa\,\rho^{\gamma_{\rm
 EOS}}$ on gas with density exceeding $n_{\rm H} = 0.1\cmcubed$, the critical
 density for the onset of the thermo-gravitational instability
 \citep{Schaye_04}. \citet{Schaye_and_Dalla_Vecchia_08} demonstrate that the
 exponent $\gamma_{\rm EOS}=4/3$ makes both the Jeans mass and the ratio of the
 SPH kernel and the Jeans length independent of the gas density, thereby
 preventing spurious fragmentation due to a lack of numerical resolution. We
 adopt this value here. The equation of state is normalised such that $P/k =
 10^3\cmcubed\,{\rm K}$ for atomic gas at the density threshold.

  Observations of galaxies reveal a tight relationship between gas
  column density and star formation rate, the Kennicutt-Schmidt law
  \citep[e.g.][]{Kennicutt_98}. \citet{Schaye_and_Dalla_Vecchia_08}
  show that since gas column density in a self-gravitating system is
  directly related to pressure, it is possible to implement a local
  Kennicutt-Schmidt law as a pressure law, independently of the chosen
  value of $\gamma_{\rm EOS}$ and even for very low numerical
  resolution. The star formation algorithm therefore converts gas
  particles with densities above the threshold $n_{\rm H} =
  0.1\cmcubed$ into stars at a rate that depends on the gas pressure, 
  \begin{equation} 
    \dot{m}_\star = A(1\Msun\pc^{-2})^{-n}m_{\rm
g}\left(\frac{\gamma}{G}f_{\rm g}P\right)^{(n-1)/2}, 
  \end{equation}
  where $m_{\rm g}$ is the mass of the gas particle for which we are
  computing $\dot{m}_\star$, $\gamma = 5/3$ for a monatomic gas (and
  should not be confused with $\gamma_{\rm EOS}$), $f_{\rm g}$ is the
  mass fraction in gas (assumed here to be unity), and $P$ is the gas
  pressure. The normalisation, $A$, and slope, $n$, follow from the
  observed \citet{Kennicutt_98} law,
\begin{equation} 
  \dot{\Sigma}_\ast = 1.5\times 10^{-4}\Msunyr\,\kpc^{-2}\left
({\Sigma_{\rm g} \over 1\Msun\,{\rm pc}^{-2}}\right )^{1.4}, 
\end{equation}
where we have divided Kennicutt's normalisation by a factor 1.65 to
account for the fact that we assume the IMF takes the form proposed by
\citet{Chabrier_03} rather than the commonly adopted form due to
\citet{Salpeter_55}.  The conversion of gas particles into stars,
based upon their associated star formation rate, is stochastic, and
since the spawning of multiple star particles from a gas particle can
lead to an artificial reduction in the efficiency of feedback
prescriptions
\citep[for a discussion, see][]{Schaye_and_Dalla_Vecchia_08}, we
convert entire gas particles into stars. Conveniently, this practice 
also conserves the particle number throughout our simulations.

\item {\em Kinetic feedback.} Feedback from star formation in the disc
  is represented in part by the imposed relation
  $P=\kappa\,\rho^{\gamma_{\rm EOS}}$. However, observed starburst
  galaxies exhibit signs of galaxy-wide winds
  \citep{Heckman_Armus_and_Miley_90,Martin_99,Pettini_et_al_02,Adelberger_et_al_03,Shapley_et_al_03,Wilman_et_al_05,Swinbank_et_al_07} 
  that are thought to be responsible for enriching the IGM with
  metals. We model the generation of winds as follows. After a delay
  of $3\times 10^7\yr$, corresponding to the maximum lifetime of
  stars that undergo core collapse SNe, newly formed star particles
  impart a randomly directed $600\kms$ kick to, on
  average, $\eta\equiv\dot m_{\rm wind}/\dot m_\star=4$ of its
  neighbours. This mass loading value was chosen to scale the global
star formation rate density, \sfrdz, such that it reasonably matches
observational data. 

 Assuming that each star with initial mass in the range
  $6-100\Msun$ injects a kinetic energy of $10^{51}\erg$, these
parameter values imply that the total wind energy 
  accounts for 80 per cent of the available kinetic energy for our
  chosen IMF. Note that contrary to the widely used kinetic feedback
  recipe of \citet{Springel_and_Hernquist_03a}, wind particles are
  injected \emph{locally} and are \emph{not} temporarily decoupled
  hydrodynamically. As discussed in
  \citet{Dalla_Vecchia_and_Schaye_08}, these changes have a large
  effect on the structure of galaxy discs and outflows.

\item {\em Chemodynamics.} Each star particle represents a single
  stellar population, with metal abundances inherited from its parent
  gas particle. Given our assumed initial mass function
  \citep[][stellar mass range $0.1 - 100\Msun$]{Chabrier_03} and
  stellar evolution tracks dependent on metal abundance
  \citep{Portinari_Chiosi_and_Bressan_98,Marigo_01,Thielemann_et_al_03_short}, we follow the
  delayed release of 11 elements (hydrogen, helium, carbon, nitrogen,
  oxygen, neon, magnesium, silicon, sulphur, calcium, and iron) by
  type Ia and type II SNe, and AGB stars. Star particles distribute
  the synthesised elements and lost mass during each timestep to
  neighbouring gas particles using the SPH interpolation
  scheme. Throughout this study, we use the solar abundances of
  \textsc{cloudy}; solar metallicity is therefore taken to be
  $\Zsun=0.0127$.
\end{itemize}

\subsection{Initial conditions and run details}
\label{sec:ics}

For brevity, we present here only an outline of the generation of the initial
conditions; a detailed technical description is deferred to the Appendix. There,
we also discuss how we address the existence of an artificial boundary between
the high-resolution region and the external, low-resolution pressureless region,
and how we combine results from the five individual \gimic\ regions to construct
estimates of properties for the entire Millennium Simulation volume.

We follow the evolution of a wide range of environments by simulating
five spherical regions whose overdensities at $z=1.5$ are
$(-2,-1,0,+1,+2)$ times the root-mean-square deviation, $\sigma$, from
the mean on some spatial scale. Ideally the simulations would resolve
the Jeans mass at our imposed star formation threshold density, but
this remains beyond the scope of our available computational
resources. Therefore, for our highest resolution simulation, we aimed
to resolve the Jeans mass in the post-reionisation IGM, which was
shown by \citet{Theuns_et_al_98} to be a prerequisite for attaining
converged properties of the Ly$\alpha$ forest. This requires a gas
particle mass of $\sim10^6\hMsun$, limiting the size of the regions to
a radius of $18\hMpc$. To ensure we also followed a rich cluster, we
imposed a precondition that the $+2\sigma$ region be centred on a
massive dark matter halo at $z=0$ and increased its radius to
$25\hMpc$.

In order to preserve the correct large-scale forces, the volume
external to these five regions was simulated at lower resolution and
without baryonic physics. This methodology is shown schematically in
Fig.~\ref{fig:VisStages}. The multiresolution particle distribution
generated for each simulation is explained in the Appendix; the
particle displacement fields were calculated using the techniques
described in \citet{Power_et_al_03} and \citet{Navarro_et_al_04_short}.

\begin{figure*}
  \includegraphics[width=\textwidth]{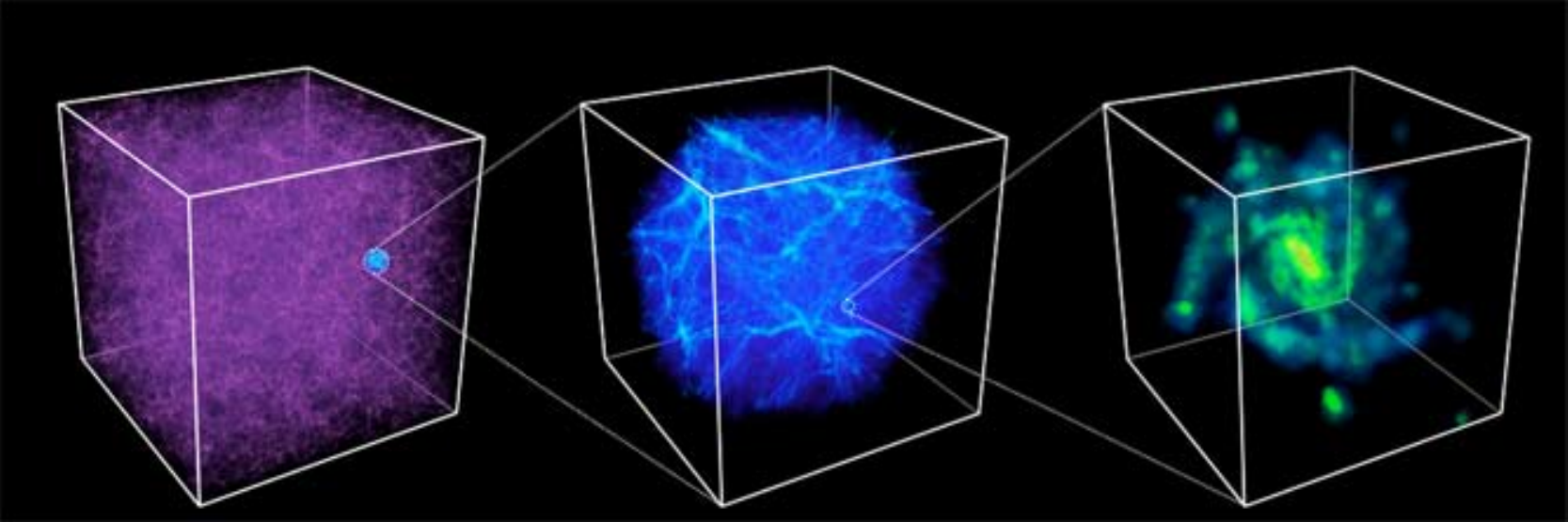}
  \caption{Schematic representation of \textsc{Gimic}, illustrating
  the large dynamic range allowed by the ``zoomed'' initial conditions
  technique. The left panel shows the density distribution within the
  $L=500\hMpc$ periodic volume of the Millennium Simulation at
  $z=1.5$. Shown embedded within this volume, to scale, is the
  $0\sigma$ \textsc{Gimic} region zoomed down to $L\simeq50\hMpc$ and
  showing the gas density. The right-hand panel, zoomed yet further to
  $L\simeq50\hkpc$, depicts the gas density within a disc galaxy of
  mass similar to that of the Milky Way.}  \label{fig:VisStages}
\end{figure*}

 \begin{table*} \begin{center} {\footnotesize
 \begin{tabular}{|l|c|c|c|c|c|c|c|} \hline\hline Region & x & y & z &
 Comoving radius & $N$ (int. res) & $N$ (high res) & $z_{\rm final}$
 \\ &[$\hMpc$] &[$\hMpc$] &[$\hMpc$] &[$\hMpc$] & & & [int. res, high
 res] \\ \hline $-2\sigma$ & 153.17 & 347.90 & 424.81 & 18 &
 $2.23\times10^{7}$ & $1.78\times10^{8}$ & [0,0] \\ $-1\sigma$ &
 387.91 & 316.48 & 113.46 & 18 & $2.80\times10^{7}$ &
 $2.24\times10^{8}$ & [0,2] \\ $\phantom{+} 0\sigma$ & 271.94 & 108.29
 & 107.45 & 18 & $3.44\times10^{7}$ & $2.75\times10^{8}$ & [0,2] \\
 $+1\sigma$ & 179.51 & 379.22 & 196.64 & 18 & $4.30\times10^{7}$ &
 $3.44\times10^{8}$ & [0,2] \\ $+ 2\sigma$ & 233.10 & 139.30 & 387.38
 & 25 & $1.24\times10^{8}$ & N/A & [0,-] \\ \hline \end{tabular} }
 \end{center} 
\caption{Parameters for the five \textsc{Gimic}
 regions. Columns 2-5 give the location (in Millennium Simulation
 coordinates) and the nominal comoving radius of the regions at
 $z=1.5$. The following two columns show the number of gas (or,
 equivalently, dark matter) particles within the zoomed region of the
 simulation, whilst the final column gives the redshift at which each
 simulation was terminated. Note that the $+2\sigma$ region was not
 run at high resolution.}  
\label{tab:sim_params}
\end{table*}

We created two realisations of the $18\hMpc$ spheres: high-resolution
(marginally resolving the IGM Jeans mass) and intermediate-resolution
(with a factor of 8 fewer particles); we reserve the term
``low-resolution'' for the original Millennium Simulation
realisation. The $+2\sigma$ region was generated at
intermediate resolution only because of its far greater computational
demands. We have hence carried out 9 simulations in total, with the
characteristics listed in Table~\ref{tab:sim_params}. All intermediate
resolution realisations have been run to $z=0$. The least
cpu-demanding region ($-2\sigma$) was also run to $z=0$ at high
resolution, whilst the three regions, $(-1,0,+1)\sigma$, were run to
$z=2$ at high resolution. Simulations of the same region at different
resolution are needed in order to check numerical convergence. In what
follows, we perform such tests at $z \ge 2$ for the $(-1,0,+1)\sigma$
and all the way to $z=0$ for the $-2\sigma$ region.

In all simulations, the gravitational forces of the baryonic and high
resolution dark matter particles were softened over the same
length scale. The softening length is initially fixed in comoving
space, but becomes fixed in physical space at a predefined redshift,
i.e. $\epsilon_{\rm com}(a)^{'} = \min(\epsilon_{\rm
com},\epsilon_{\rm phys}^{\rm max}/a)$. The softenings were chosen
such that at $z=3$, they are fixed at $\epsilon_{\rm phys}^{\rm max} =
(1.0,0.5)\hkpc$ for the intermediate- and high-resolution runs,
respectively. The gas particle masses are $1.16\times 10^{7}~\hMsun$ and $1.45\times
10^{6}~\hMsun$ at intermediate and high resolution
respectively. These are much smaller than the limit required to avoid
artificial two-body heating effects
\citep{Steinmetz_and_White_97} that dominate over radiative cooling. The dark particle masses are a
factor $(\Omega_{\rm m}-\Omega_{\rm b})/\Omega_{\rm b}=4.56$ larger that the gas particle
masses. 

\subsection{Halo and galaxy identification}
\label{sec:halo_and_galaxy_identification}

We identify haloes by applying the friends-of-friends (FoF) algorithm to dark matter
particles, using the standard value of the linking length in units of
the mean interparticle separation \citep[$b=0.2;$][]{DEFW85}. In order
to identify the baryonic content of dark matter haloes, our 
group finding algorithm locates the nearest dark matter particle to
each baryonic (i.e. gas or star) particle. If this dark matter
particle has been grouped by FoF, the corresponding baryonic particle
is also associated with the FoF group. The assignment of baryons to
dark matter haloes is therefore unambiguous. This method occasionally
introduces artefacts, mostly for groups with few particles. For
example, we find that the baryon fraction of low-mass haloes varies
widely, from haloes with almost no baryons, to haloes with more than
the cosmic mean value, because our scheme allows the baryon
distribution around a halo to be more extended than the dark
matter. Since these problems are mostly restricted to small haloes
close to the resolution limit, we have not attempted to overcome
them. 

The FoF algorithm identifies isodensity contours of $\delta
\simeq 3/(2\pi b^3) \simeq 60$ \citep{Lacey_and_Cole_94}, but noise in
the particle distribution leads to the detection of many small,
artificial objects clustered around the true halo. Since these
artefacts tend to be transient, we perform an unbinding calculation
using the \textsc{Subfind} algorithm
\citep{Springel_et_al_01,Dolag_et_al_08}, and omit from our analyses
any FoF haloes that do not have at least one self-bound
substructure. This additional procedure successfully removes
artificial haloes from our sample. The version of \textsc{Subfind}
used here is modified from the standard implementation, such that
baryonic particles are also considered when identifying self-bound
substructures \citep[for more details see][]{Dolag_et_al_08}. This
provides an unambiguous definition of a \textit{galaxy} within the
simulations as the set of star particles bound to individual
substructures. Thus, an individual halo may host more than one galaxy.

When quoting halo masses, we refer to the total mass of the FoF halo
in all components (i.e. gas, stars and cold dark matter). In some
cases, however, it is appropriate to modify this definition
slightly. When comparing results with simulations that follow only
dark matter (e.g.~Section \ref{sec:haloes}), we consider only the cold
dark matter component of a FoF halo and boost its mass by a factor of
$\Omega_{\rm m}/(\Omega_{\rm m}-\Omega_{\rm b})$ to account for the
baryonic component. Additionally, when measuring baryon fractions it
is common to specify fractions within the well-defined volume of a
spherical overdensity (SO) group \citep[e.g.][]{Lacey_and_Cole_94}; in
this case we quote the mass within a sphere, centred on the local
minimum of the gravitational potential, whose radius encloses a mean
density (in all components) of $200\rho_{\rm c}$, where $rho_{\rm c}$ is the critical density for a flat Universe. We also quote the
circular velocity, $v_{\rm c}$, of haloes at this radius, $r_{200}$.

\section{Evolution of the halo and stellar mass functions}
\label{sec:haloes}

\begin{figure*}
  \includegraphics[width=0.48\textwidth]{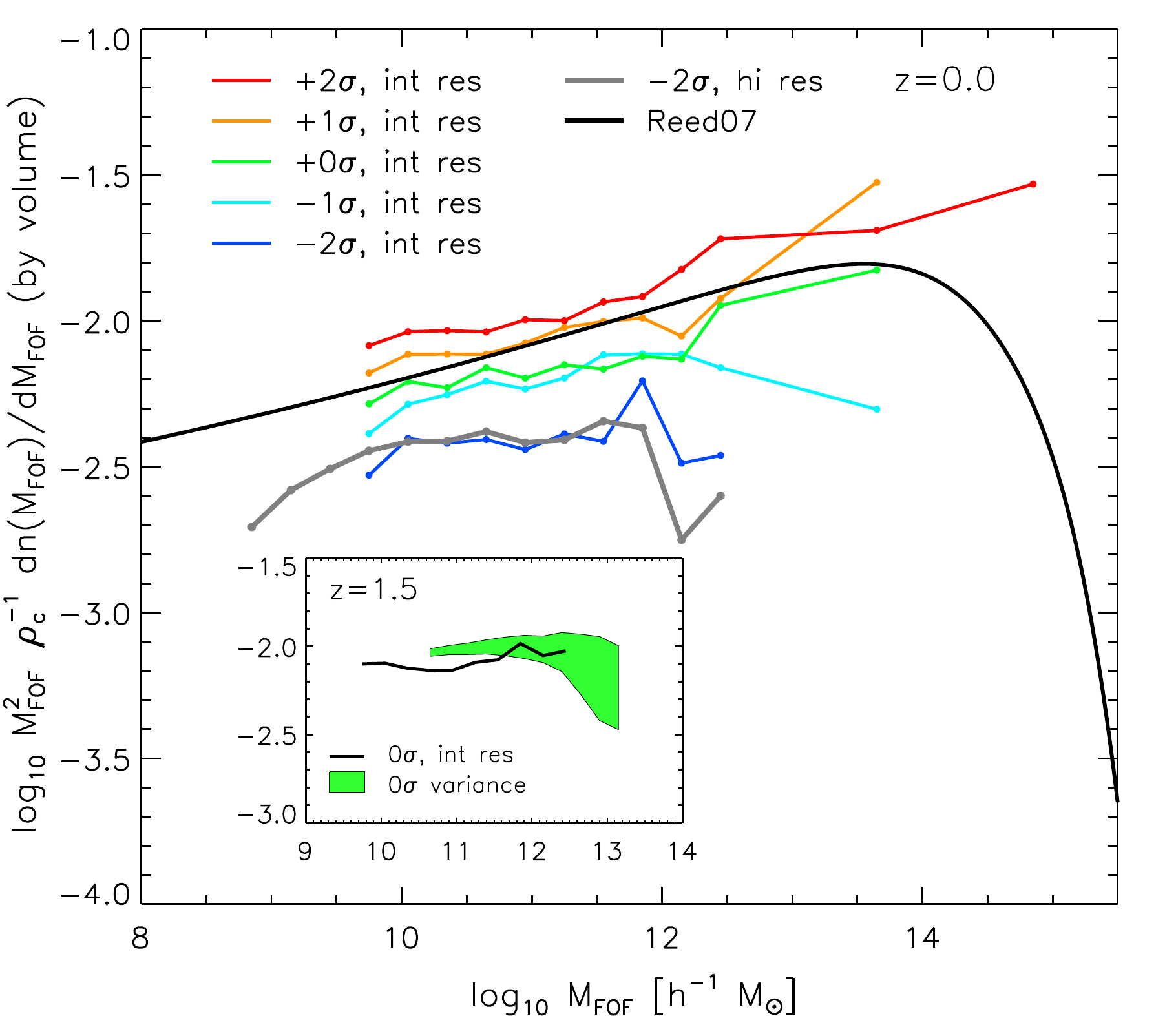}
  \includegraphics[width=0.48\textwidth]{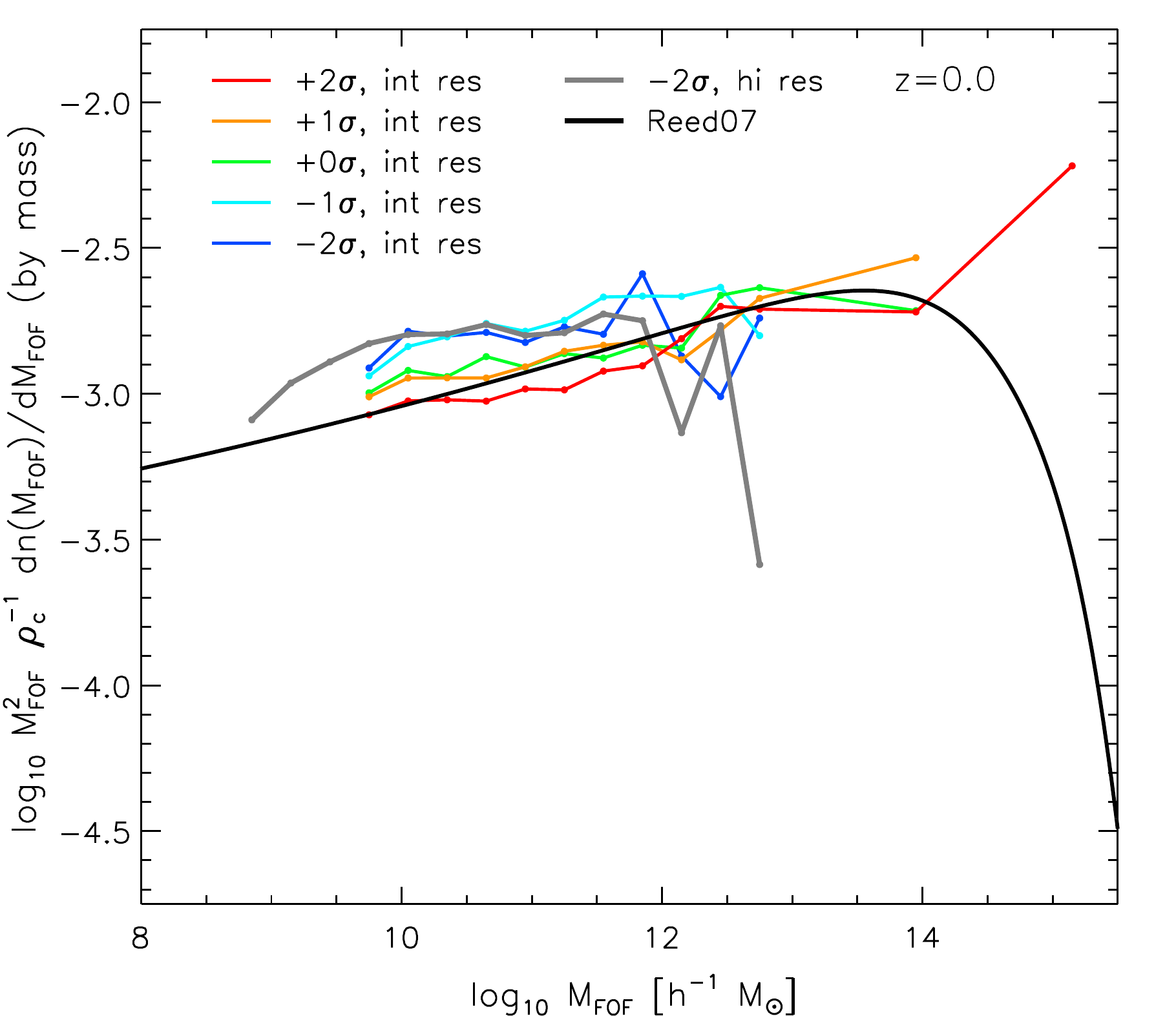}
  \caption{\textit{Left:} Differential number density of haloes as a
  function of mass at $z=0$, normalised by volume, and multiplied by
  $M^2$ in order to reduce the dynamic range of the plot. Coloured
  lines represent the intermediate-resolution \gimic\ regions, and the
  grey line represents the $-2\sigma$ region at high resolution. The
  global halo mass function of \citet[][\textit{black}]{Reed_et_al_07}
  is bracketed by the extreme \gimic\ regions. Inset is the halo mass
  function at $z=1.5$, plotted for the $0\sigma$ region. The green
  shaded area delineates the $16^{\rm th}$ and $84^{\rm th}$
  percentiles of the mass functions of 100 candidate $0\sigma$ spheres
  drawn from the Millennium Simulation. \textit{Right:} As left, but
  the differential number density is normalised by the total enclosed
  mass of each region, rather than by its volume. The difference
  between the regions is smaller when the regions are normalised by
mass, but there is still a systematic variation from region to region
at all masses. The most massive haloes form only 
in the most overdense, and hence most dynamically advanced, regions.}
  \label{fig:halo_mass_function}
\end{figure*}

\subsection{The halo mass function in different environments}

As pointed out by \citet{Frenk_et_al_88}, and discussed more recently
by \citet{Cole_97} and \citet{Sheth_and_Tormen_02}, a fundamental
difference between regions of varying overdensity is the rate at which
their dark matter halo populations evolve. This is reflected in the
clustering bias of haloes as a function of mass
\citep{Mo_and_White_96}. As we shall see below, many environmental effects
revealed by our simulations, for example, variations in the star
formation rate density, can be traced back to differences in the halo
mass function in different environments. We therefore preface the
discussion of star formation with a brief overview of the evolution of
the halo mass function in each region, including resolution tests.

In Fig.~\ref{fig:halo_mass_function}, we show the differential number
density of haloes identified with the FoF algorithm, $dn(M)/dM$,
multiplied by $M^2$ in order to reduce the dynamic range of the plot,
and thus more clearly highlight the differences between regions. As
noted in Section~\ref{sec:halo_and_galaxy_identification}, the halo
mass in this plot is taken to be $M_{\rm FOF} = M_{\rm FOF}^{\rm
DM}\Omega_{\rm m}/(\Omega_{\rm m}-\Omega_{\rm b})$, so as to enable
the most direct comparison possible with the results of dark matter
only simulations. The regions simulated at intermediate resolution are
shown as coloured curves, while the high-resolution realisation of the
$-2\sigma$ region is shown as a grey curve. In the region of overlap,
the agreement between the high- and intermediate-resolution
realisations is excellent, suggesting that the mass function in the
latter has converged down to $1\times 10^{10}\hMsun$ (corresponding to
$\sim 200$ particles). Extrapolating to the high-resolution
simulation, we estimate that the halo mass function is reliable down
to $\sim 10^9\hMsun$. For subsequent analyses of the baryonic
properties of haloes, we therefore consider only FoF haloes with at
least 200 particles.

The \gimic\ mass functions are also compared with the global mass
function fit of \citet{Reed_et_al_07}, which extends to both higher
and lower mass scales since it was derived from an ensemble of dark
matter simulations covering a wide dynamic range in mass
\citep[including the Millennium Simulation; see
also][]{Jenkins_et_al_01}.  As expected, the mass functions of the
under- and overdense \gimic\ regions bracket the global function,
which is reasonably well matched by the mass functions of the $0$ and $+1\sigma$
regions.  We have checked that the halo mass functions of the 
regions at $z=1.5$ (when they were selected) are consistent with those
of all possible candidate spheres in the Millennium Simulation. As an
example, the inset shows the multiplicity function of the
intermediate-resolution $0\sigma$ region, overplotted on the locus
that encompasses the $16^{\rm th}$ and $84^{\rm th}$ percentiles of
the multiplicity functions of 100 spheres, drawn from the Millennium
Simulation, that qualified as candidates for the $0\sigma$ region (for
details see the Appendix).

Although the different halo mass functions appear as scaled versions
of one another at high redshift, a large difference develops over
time, as the amplitude and shape of the function evolves. By the
present day, there are approximately 3 times more haloes of
intermediate mass in the $+2\sigma$ region compared to the $-2\sigma$;
the difference increases with mass, with large mass haloes found only
in the $+2\sigma$ region. To determine the efficiency of halo
formation per unit mass, we show in the right-hand panel of
Fig.~\ref{fig:halo_mass_function} the halo mass function now
normalised by the total enclosed mass, rather than by the total
enclosed volume. Normalised this way, the differences between the
\gimic\ regions are smaller but there is still a large systematic
variation from region to region at all masses. Haloes form more
efficiently in the high-density regions because they are the most
dynamically advanced and the most massive haloes form only in the most
overdense regions. These results agree with previous numerical and
analytical studies of dark matter halo evolution
\citep[e.g.][]{Frenk_et_al_88,Mo_and_White_96,Sheth_and_Tormen_02}. 

\subsection{The galaxy stellar mass function}

\begin{figure*}
  \includegraphics[width=0.49\textwidth]{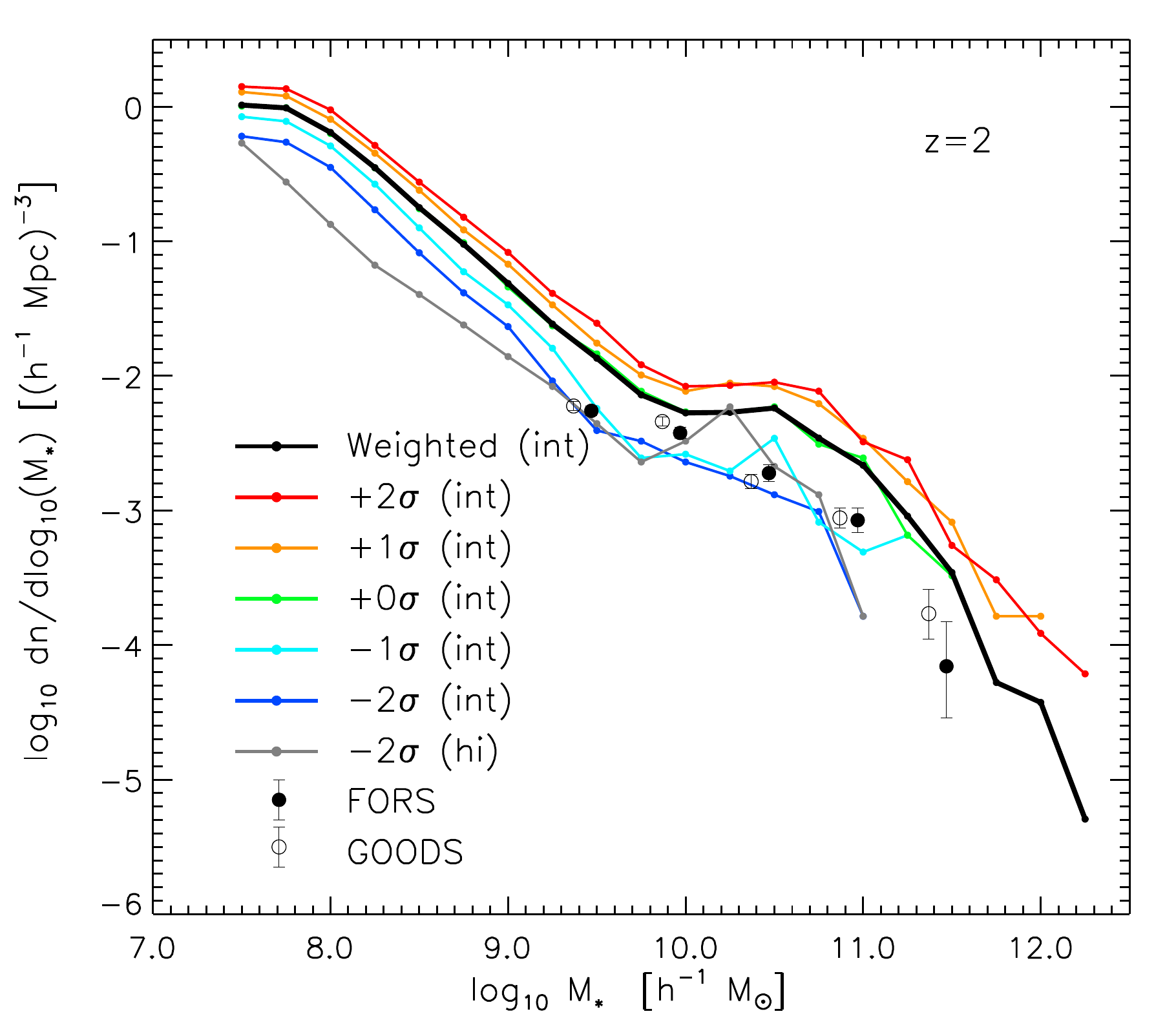}
  \includegraphics[width=0.49\textwidth]{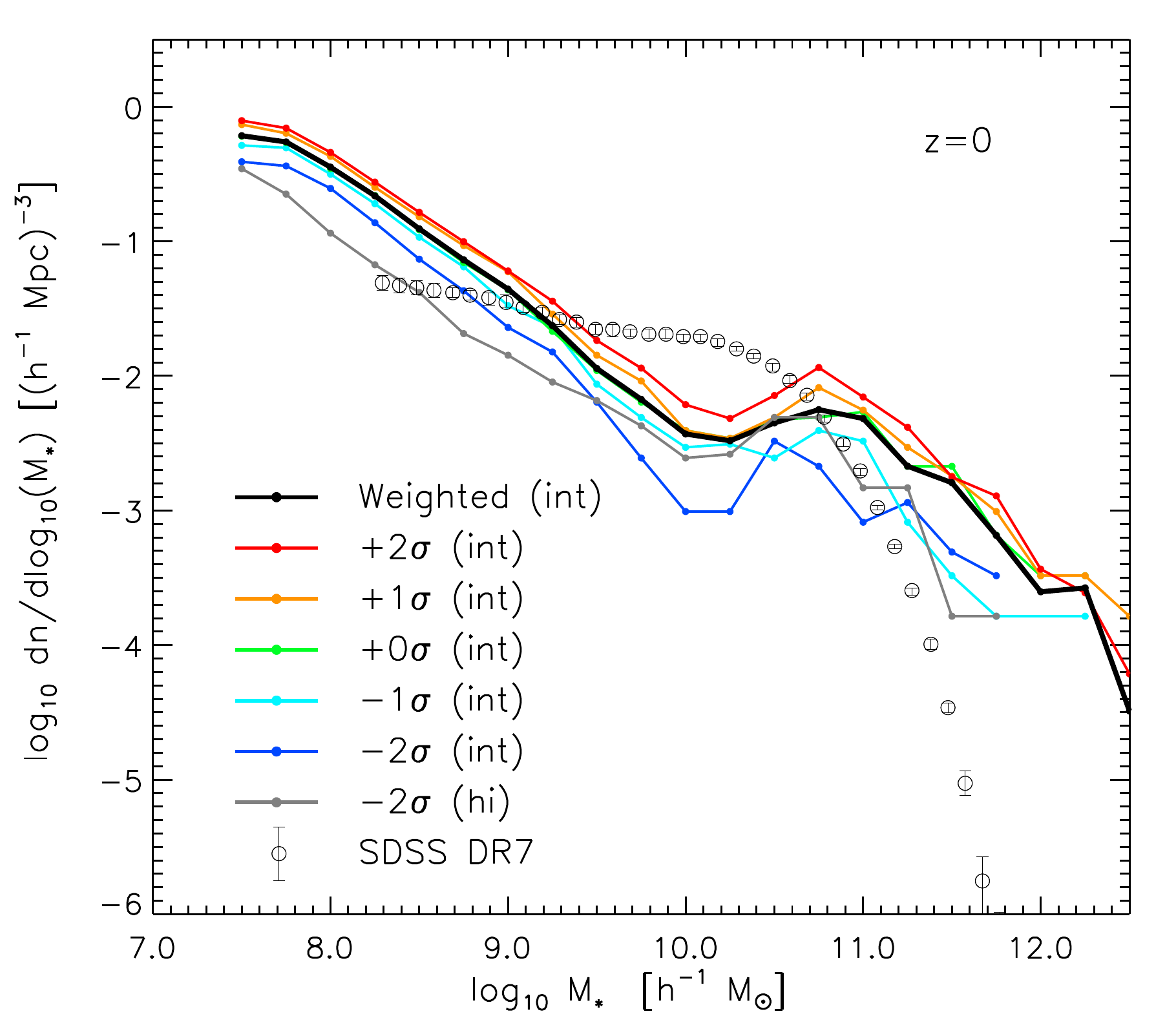}
  \caption{The stellar mass function of galaxies at $z=2$
  (\textit{left}) and $z=0$ (\textit{right}). Results are shown for
  all five intermediate-resolution simulations (\textit{coloured
  curves}), and their weighted average (\textit{black curve}), which
  represents an estimate of the function over the entire Millennium
  Simulation volume. The stellar mass function of the $-2\sigma$
  region at high resolution is also shown (\textit{grey curve}) to
  illustrate the degree of convergence. Observational data from the
  FORS and GOODS deep fields \citep{Drory_et_al_05} are overplotted at
  $z=2$, and from SDSS DR7 \citep{Li_and_White_09} at $z=0$. The
  simulations are in rough agreement with the data at $z=2$; at
  $z=0$ the overall number of galaxies more massive than $\sim
  10^9\Msun$ is correctly reproduced by the simulations, but the
  number density of intermediate mass galaxies is too low, whilst that
  of the most massive galaxies is too high. }
\label{fig:stellar_mass_function}
\end{figure*}

The galaxy stellar mass function as a function of redshift provides a
useful check of the realism of our simulations. This function is shown
in Fig.~\ref{fig:stellar_mass_function} and compared with
observational data at $z=2$ and $z=0$. The stellar mass functions in
each of the five \gimic\ regions at intermediate resolution are shown
as coloured curves. From these, we obtain an estimate of the
stellar mass function for the entire Millennium Simulation volume
(\textit{black curve}), by a suitably weighted average; details of
this procedure are given in the Appendix.

As in Fig.~\ref{fig:halo_mass_function}, we also plot results from the
high-resolution realisation of the $-2\sigma$ region (\textit{grey
curve}) as a numerical test. This shows that at $z=2$, the
intermediate-resolution stellar mass function has approximately
converged for $M_\star \ge 10^9 \hMsun$. Below this value, there are
roughly twice as many galaxies in the intermediate-resolution
simulation. This excess almost certainly reflects a reduction in the
efficiency of supernovae feedback in poorly resolved galaxies. At
$z=0$, the convergence properties are similar: below $M_\star \sim
10^9 \hMsun$, the intermediate-resolution simulation overpredicts the
number of galaxies by about a factor of 2, and above this mass, it
underpredicts it by a similar factor.

The weighted average stellar mass function closely tracks that of the
$0\sigma$ region over the mass range for which haloes are present in
this region. At the high mass end, only the overdense regions
contribute to the stellar mass function. This highlights the need to
follow a wide range of large-scale environments when attempting to
compare simulations with large galaxy redshift surveys. For
comparison, we overplot the stellar mass functions derived from the
FORS and GOODS deep fields at $z=2$ \citep{Drory_et_al_05} and from
SDSS data at $z=0$ \citep{Li_and_White_09}.  At $z=2$, the data span a
factor of 100 in stellar mass. Over this limited range, the model
stellar mass function has a similar shape to the data but the
amplitude of the weighted model function is approximately 0.5~dex
higher, which is a slightly less than the difference between this
model function and that of the $-2\sigma$ region.

At $z=0$, the data span nearly 4 orders of magnitude in stellar
mass. The simulations produce too many small galaxies, with $M_\star<10^9
\Msun$, and too few in the mass range $10^9-10^{10}\Msun$, but
inspection of the cumulative function shows that they produce about
the right number density at $M_\star> 10^{9}\Msun$ and then at $M_\star>2.5 \times
10^{10}\Msun$. At larger masses, the two overdense \gimic\ regions
produce a population of massive galaxies that is not seen in the
data. The overproduction of faint galaxies and the ``dip'' at
intermediate masses result from the simple wind model that we have
adopted. The excess at large masses in the overdense regions results
from our neglect of heating processes that can quench cooling flows in
clusters. In semi-analytic models of galaxy formation \citep{Bower_et_al_06,Croton_et_al_06_short,Somerville_et_al_08} and cosmological hydrodynamical simulations \citep{Sijacki_et_al_08,DiMatteo_et_al_08,Booth_and_Schaye_09} this is achieved
by invoking feedback from AGN, which is not modelled in our simulations.

The sensitivity of the stellar mass function to variations in the
physical assumptions and parameters in the simulations will be
explored in the OWLS project (Haas et al.,
\textit{in prep}). For our purposes, it is sufficient to
note that there is reasonably good agreement between the simulations
and the data at high redshift while, at low redshift, the simulations
produce roughly the right number of galaxies, albeit with an incorrect
distribution of stellar masses.

\section{The evolution of cosmic star formation}
\label{sec:sfrd}

\begin{figure*}
  \includegraphics[width=0.49\textwidth]{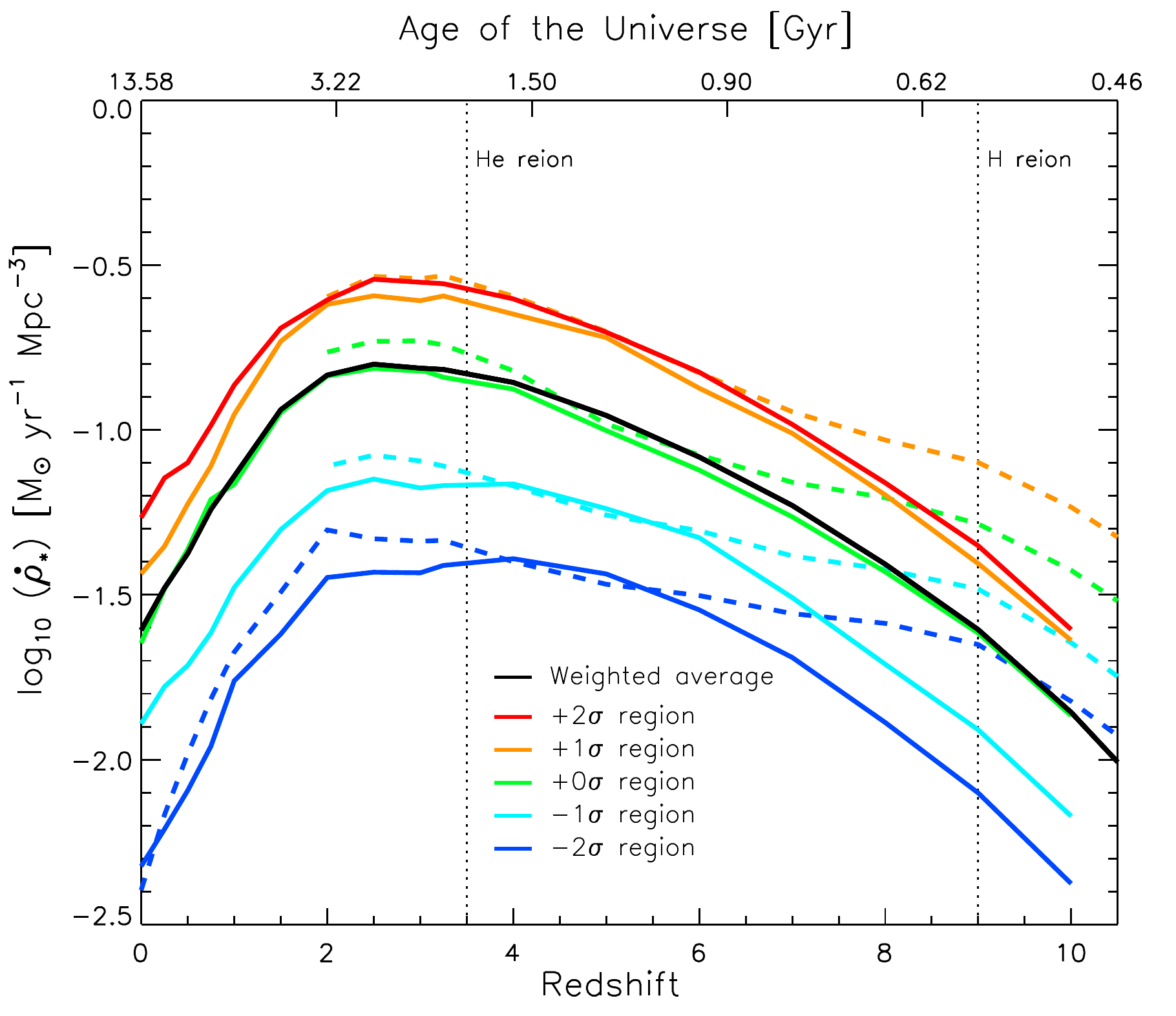}
  \includegraphics[width=0.49\textwidth]{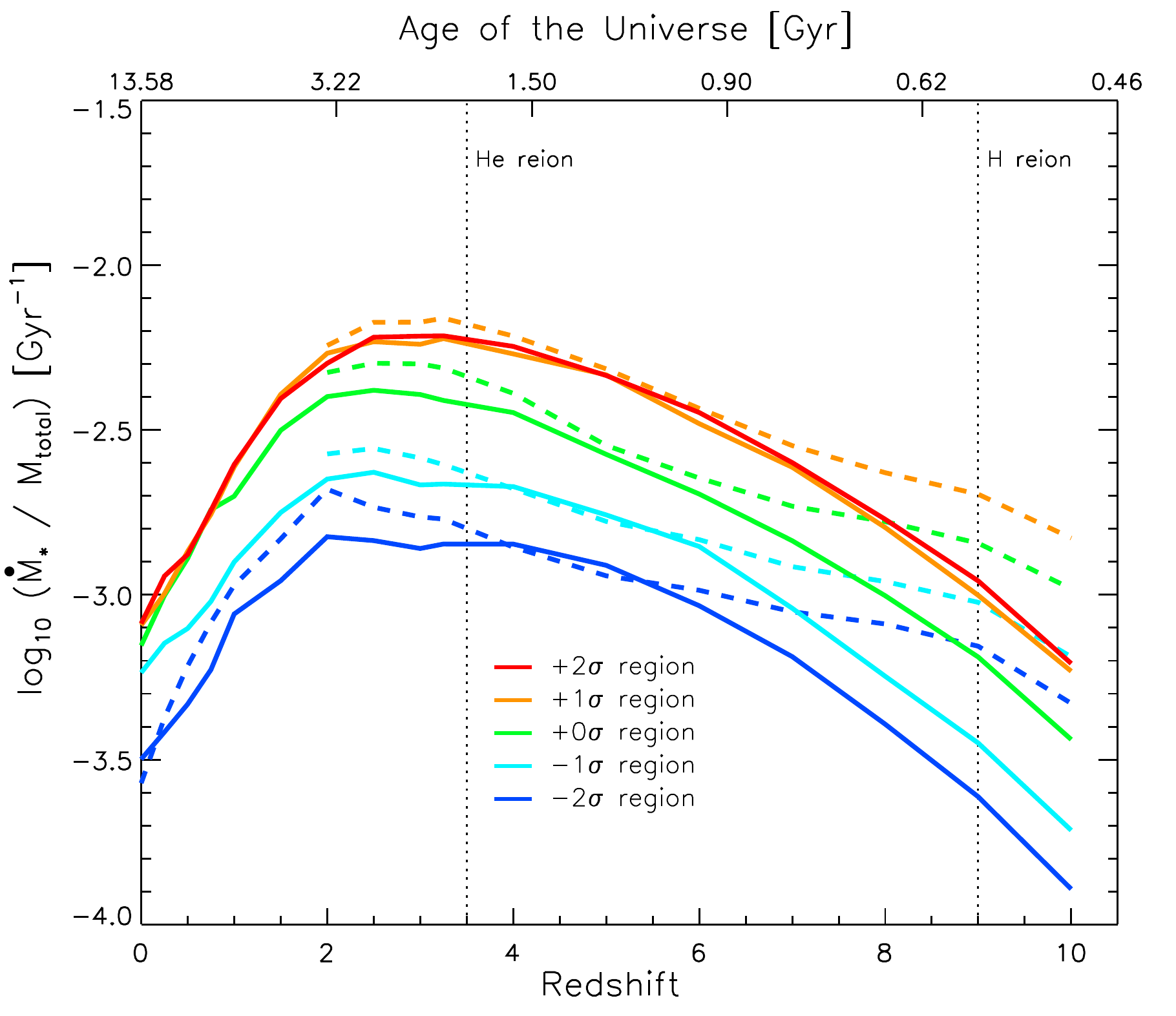}
  \caption{ \textit{Left:} Star formation rate density,
  $\dot{\rho}_\star(z)$, as a function of redshift in regions of
  different overdensity (\textit{colours}), at intermediate (\textit{solid lines}) and high resolution (\textit{dashed
  lines}). We have not run the $+2\sigma$ simulation at high
  resolution. The vertical dotted lines denote the epochs of
  reionisation of hydrogen (H\textsc{i}) and helium (He\textsc{ii})
  respectively.  The variation in $\dot{\rho}_\star$ amongst the
  large-scale environments is pronounced; the most extreme regions
  differ by an order of magnitude at all epochs.  \textit{Right:} The
  evolution of the total SFR {\em per unit mass}, in regions of
  different large-scale overdensity. The persistent offsets demonstrate
  that the increase of $\dot{\rho}_\star$ with large-scale overdensity does
  not arise purely because overdense regions enclose a greater mass.}
  \label{fig:Madau_CONVERGENCE}
\end{figure*}

The cosmic star formation rate density (SFRD) is a key tracer  of
structure formation in the universe. After more than a decade of
observational studies, it appears to be fairly well constrained at
least out to intermediate redshifts ($ z \lesssim 3$), given
assumptions about the initial mass function and reddening
\citep{Hopkins_07}. Taking
into account the mass fraction recycled during stellar evolution, an
integral over \sfrdz\ gives the present day cosmic stellar mass
density
\citep[e.g.][]{Cole_et_al_01_short,Rudnick_et_al_03_short,Eke_et_al_05,Li_and_White_09}.
These are all quantities that are calculated in our simulations.

In this section we investigate the star formation history in our
simulations and how it varies from one
\gimic\ region to another. We discuss the numerical convergence of our
results and the nature of the dominant contributors to the star
formation activity at high-redshift. The averaging technique described
in the Appendix is applied to construct an estimate of the global SFRD
that can be compared both to observations and to the results of the
semi-analytic calculations applied to the Millennium Simulation.

\subsection{Large-scale environmental modulation}
\label{sec:large-scale variation}

\begin{figure*}
  \includegraphics[width=0.49\textwidth]{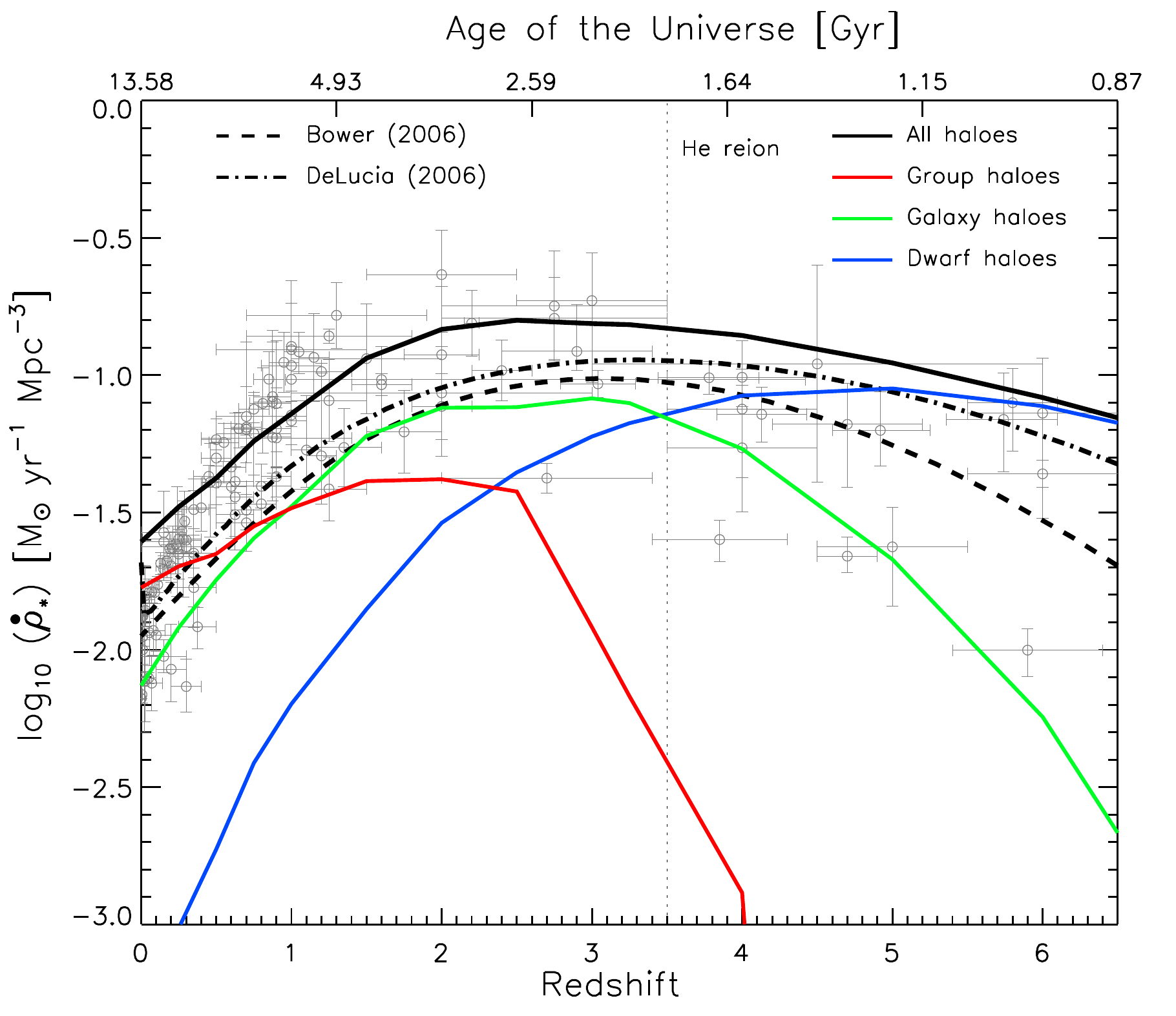}
  \includegraphics[width=0.49\textwidth]{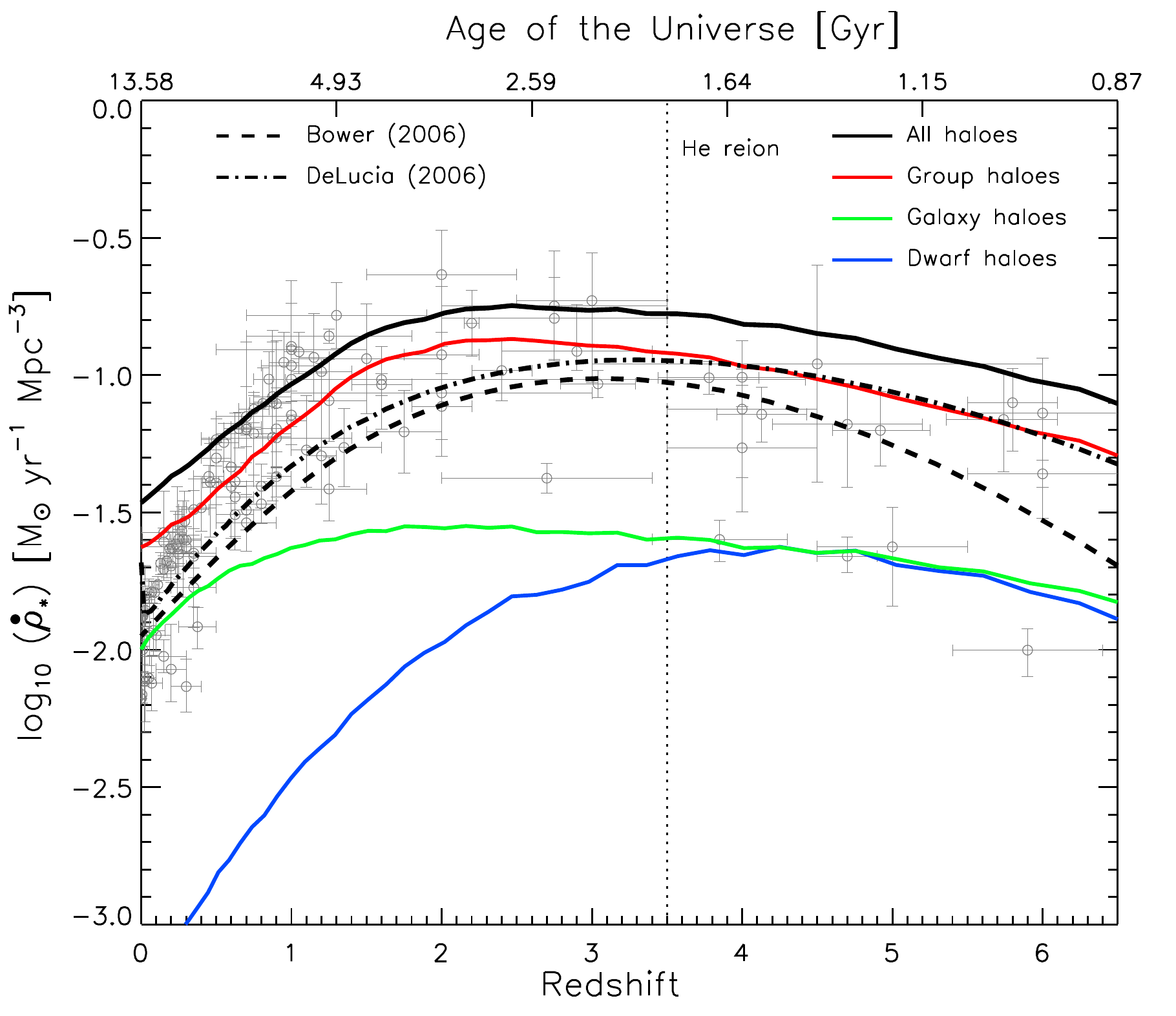}
  \caption{\textit{Left:} Star formation rate density,
    $\dot\rho_\star(z)$, in the $L = 500\hMpc$ Millennium
    Simulation volume (\textit{black solid line}), derived from a
    weighted average of the intermediate-resolution
    \gimic\ simulations. Coloured lines show the instantaneous
    contribution to the total from dwarf galaxy haloes
(\textit{blue}), regular galaxy haloes (\textit{green}) and
group/cluster 
    haloes (\textit{red}).  The evolution of \sfrd\ predicted by the
    semi-analytic calculations of \citet[][\textit{black dashed
        line}]{Bower_et_al_06} and \citet[][\textit{black dot-dashed
        line}]{De_Lucia_et_al_06} are shown for comparison. Symbols
    with error bars are observational estimates from the compilation of
    \citet{Hopkins_07}, converted to the Chabrier IMF. \textit{Right:}
    The ``archaeological'' description of the star formation rate density. Haloes are assigned to 
    bins according to their mass at $z=0$ and the coloured curves
    show the value of \sfrdz\ \textit{due to their progenitors}.}
\label{fig:Madau_HaloMassSplit}
\end{figure*}

For each region we consider all gas particles within the sampling
sphere discussed in the Appendix, and use the volume of the sphere to
compute the star formation rate per unit volume,
$\dot{\rho}_\star(z)$, shown in
Fig.~\ref{fig:Madau_CONVERGENCE}. Comparison of the intermediate- and
high-resolution simulations shows that our results have not converged
at $z \gtrsim 6$; the high-resolution realisations exhibit significantly
higher overall star formation rates than their intermediate-resolution
counterparts. At later times, however, convergence of \sfrdz\ is
achieved to within $\sim 50$~per cent, and inspection of 
the curves for the $-2\sigma$ region realisations demonstrates that
this conclusion extends to the present epoch.

The evolution of \sfrdz\ may depend on resolution because the
simulation fails to form small mass haloes, and/or because the star
formation rate in the haloes that do form depends on resolution. The
particle mass of the high-resolution realisations was chosen so that
the Jeans mass in the photoionised IGM is marginally resolved. Hence
these simulations should be able to follow the formation of all haloes
in which the gas cools by atomic line cooling after reionisation. We
show later that the star formation rate for resolved haloes is very
similar at the two resolutions. We are therefore reasonably confident
that our estimate of $\dot{\rho}_\star(z)$ is close to numerically
converged in our high-resolution simulations for $z < z_{\rm reion} =
9$, but clearly this is not the case in the intermediate-resolution
realisations for $z\gtrsim 6$. The actual value of the star formation
rate is, of course, sensitive to our subgrid modelling, in particular
to the implementation of galactic winds. Note that the lack of
high-redshift agreement in the value of
\sfrd\ between the two resolutions has only a small effect on the
total amount of stars formed and on the distribution of stellar ages at
low-redshift, because the duration of the high redshift period before
the values converge is small compared to the age of today's universe.

The main features of the star formation rate density in our simulations are:
\begin{itemize}
\item \sfrd\ increases with decreasing redshift, peaks
  at $z\sim2-3$, and then drops rapidly by a factor $\sim 6$ to $z=0$;
\item the shape of \sfrdz\ is similar for all regions;
\item the amplitude of \sfrdz\ varies by {\em an order of
  magnitude} between the most extreme regions, at all epochs.
\end{itemize}
The amplitude increases monotonically from the $-2\sigma$ to the $+2\sigma$
region at all epochs, but the difference between the $+1\sigma$ and the
$+2\sigma$ regions is far smaller than the difference between the
$-2\sigma$ and the $-1\sigma$ regions. The large variations in the
amplitude of \sfrdz\ between the different regions is not
simply caused by the greater mass contained in the higher density
volumes. As shown in the right hand panel of
Fig.~\ref{fig:Madau_CONVERGENCE}, higher $\sigma$ regions also have a 
higher {\em specific} star formation rate, $\dot M_{\rm star}/M_{\rm
total}$ (where $M_{\rm total}$ is the total mass enclosed in each
sphere). The SFR {\em per unit mass} still varies by $\sim0.5~$dex
between the regions. We show later in Section \ref{sec:multiplicity}
that this is due to the different halo mass functions 
in regions of different overdensity, and the variation in star formation efficiency with halo mass.

A notable feature of the SFR in the high-resolution simulations
(\textit{dashed lines}) is the kink in $\dot\rho_\star$ at $z \simeq
9$, the epoch when the IGM is reionised and heated. The rapid increase
in $\dot\rho_\star$ prior to this epoch is slowed by reionisation in
all regions. The onset of reionisation has two main effects on the
evolution of gas: (i) the photoheating boosts and maintains the gas
temperature at a near-uniform level of $\sim10^4~$K, and (ii) the
photoionisation suppresses cooling due to line emission from hydrogen
and helium \citep[]{Efstathiou_92} as well as heavy elements
\citep[]{Wiersma_Schaye_and_Smith_09}. Small haloes, whose virial
temperature is below the temperature of the photoionised IGM, cannot
hold on to their photoheated gas at $z<z_{\rm reion}$, and star
formation within them ceases \citep{Pawlik_and_Schaye_09}. Both \citet{Hoeft_et_al_06} and \citet{Okamoto_Gao_and_Theuns_08} used numerical simulations to construct a simple analytic model for how the
characteristic mass, $M_c(z)$, below which haloes lose most of their
baryons, depends on the shape of the temperature-density relation for
the IGM. This characteristic mass is not well resolved in the
intermediate-resolution simulations (\textit{solid lines}) for the
ionising background adopted here \citep{Haardt_and_Madau_01}, which
explains the absence of the kink in that simulation. In accord with
the star formation history shown in the left-hand panel of
Fig.~\ref{fig:Madau_HaloMassSplit}, this also 
indicates that star formation at this epoch is dominated by low-mass
haloes ($T_{\rm vir} \lesssim 10^4 ~$K); otherwise reionisation would
have had little effect on the \textit{global} star formation rate
density.

The maximum in the star formation rate density at $z\sim 2-3$ occurs
significantly later than in the simulations of
\citet{Springel_and_Hernquist_03b}, which have comparable numerical
resolution to ours (and only slightly different cosmological
parameters; they adopted $\Omega_{\rm m} = 0.3$, $\Omega_\Lambda =
0.7$, $\Omega_{\rm b} = 0.04$, $\sigma_8 = 0.9$, $h = 0.7$). We
believe that this is only partly due to our inclusion of metal line
cooling that affects the SFR particularly at lower redshift, as
discussed and illustrated in the analytical model of
\citet{Hernquist_and_Springel_03}. The largest simulations of
\citet{Oppenheimer_and_Dave_06} cover volumes of $(32\hMpc)^3$ and
have $256^3$ particles, which results in a factor of $\sim2$ poorer mass
resolution than our intermediate-resolution simulations, and a factor $\sim15$
poorer resolution than the high-resolution \gimic\ runs. At this
level of resolution, our simulations would certainly be far from converged. The
SFR in most of their models also peaks at a higher redshift than
ours. The simulations of both \citet{Springel_and_Hernquist_03b} and
\citet{Oppenheimer_and_Dave_06} use the multiphase ISM implementation
of \citet{Springel_and_Hernquist_03a} for gas in galaxies, as well as
a different prescription for the generation of galactic winds. The
details of the ISM and wind implementations are probably the main
cause of the formation history differences between these simulations
and those presented here.

\subsection{The Millennium Simulation star formation rate density:
comparison with observations and semi-analytic models} 
\label{sec:mill_sfrd}

The star formation rates in the individual \gimic\ regions can be
combined in the manner discussed in the Appendix in order to 
estimate the net star formation rate of the whole  Millennium
Simulation volume. In order to obtain results down to $z=0$, we use the
intermediate-resolution simulations; we consider only epochs $z<6$
where our results are approximately converged. 

Our estimate of \sfrdz\ for the Millennium Simulation volume is shown in Fig.
\ref{fig:Madau_HaloMassSplit}. It increases with time to
reach a broad plateau at $\approx 0.25\MsunyrMpccubed$ over the
interval $3 \gtrsim z \gtrsim 1$ with a maximum around $z\sim 2$,
followed by a rapid decline by a factor of $\sim 6$ to $z=0$. The
behaviour of the global \sfrdz\ closely follows that of the mean
density ($0\sigma$) region.

The coloured curves in the left-hand panel decompose \sfrdz\ into
contributions from haloes binned according to their mass at that epoch;
bins are chosen to correspond approximately to the masses of haloes
that host dwarf galaxies (``dwarf haloes''; $M < 5\times
10^{11}\hMsun$), typical galaxies (``galaxy
haloes''; $5\times 10^{11}\hMsun < M < 5\times
10^{12}\hMsun$) and groups and clusters of galaxies
(``group haloes''; $M > 5\times 10^{12}\hMsun$). As might
be expected within the CDM hierarchical assembly model, the peak of
star formation within low mass haloes precedes that of more massive
haloes. For our particular classification, these peaks are broad,
occurring at $z\sim4-5$ for dwarf haloes, $z\sim2-3$ for galaxy haloes,
and $z\sim1-2$ for group haloes. 

The simulation outputs record the {\em formation time} of each star
particle and this allows us to relate present day systems to the
properties of the progenitors in which their stars formed. This
``archeological'' description of the star formation history is computed
for haloes of a given mass at $z=0$ by summing the initial mass of their
stellar particles (i.e. the mass at the time of formation, prior to mass loss due to stellar evolution), binned by their formation redshift, and is shown in
the right-hand panel of Fig.~\ref{fig:Madau_HaloMassSplit}. Note that
the solid black curve, integrating over haloes of all masses, differs
slightly from that of the left-hand plot at epochs other than $z=0$,
because the population of baryonic particles within the sphere used to
define the volume limited ``sample'' for these plots (see the Appendix
for details of how this is selected) evolves slightly over time. This
plot demonstrates that star formation is dominated at all redshifts by
the progenitors of galaxies that today reside in the most massive
haloes (i.e. groups and clusters). This is perfectly compatible with a
hierarchical build-up, because the star formation in massive haloes
became dominant only recently, as shown in the left-hand panel of the
figure.

To compare the \gimic\ simulations to observations, we use the
compilation of star formation rates given by \citet{Hopkins_07}. For
consistency, we have adjusted the published values to account for the
cosmological parameters and IMF adopted in the simulations; to convert
from the IMF of \citet{Salpeter_55} to that of \citet{Chabrier_03},
\sfrd\ is scaled down by a factor of 1.65. As noted in Section
\ref{sec:code}, the mass loading in the galactic winds 
was chosen by requiring small test simulations to produce an
approximate match to the amplitude of the observed \sfrdz. As a
consequence, the broad agreement with the data is largely a check that
the weighting scheme works well.

The evolution of the star formation rate density shows a broad plateau
of nearly constant star formation rate during $4\gtrsim z\gtrsim 1$,
followed by a rapid drop to $z=0$. The reduction in \sfrd\ towards the
present day is smaller in the simulation than in the data. This is
mostly due to the high star formation rates in massive haloes in the
simulations (red curve in the left-hand panel of
Fig.~\ref{fig:Madau_HaloMassSplit}), which is also the cause for the
overproduction of bright galaxies seen in
Fig.~\ref{fig:stellar_mass_function}. Recent semi-analytic models
\citep[e.g.][]{Bower_et_al_06,Croton_et_al_06_short,De_Lucia_et_al_06,Font_et_al_08_short} and cosmological hydrodynamical simulations \citep{Sijacki_et_al_08,DiMatteo_et_al_08,Booth_and_Schaye_09} include AGN radio-mode feedback to suppress the cooling of gas in such
haloes and quench their star formation. As a result, \sfrd\ in these
models falls faster from $z=1$ to $z=0$, in better agreement with the
data (dashed and dot-dashed lines in
Fig.~\ref{fig:Madau_HaloMassSplit}). \citet{Kobayashi_Springel_and_White_07}
obtained a qualitatively similar star formation history to that of
\gimic\ in hydrodynamic simulations with very strong feedback from
hypernovae (HNe), but no AGN feedback. Like us, they found star
formation at late times to be dominated by massive blue galaxies.

The discrepancy in \sfrd\ occurs at late times when the star formation
rate in all objects is relatively low, and therefore has little effect
on the global stellar mass density, $\Omega_\star$ (in units of the
critical density). The simulation produces a \textit{current} stellar
mass density (obtained by integrating \sfrdz\ and subtracting the mass
recycled by stellar evolution) of $\Omega_{\star}\sim 3.2\times
10^{-3}$ at $z=0$. Observational estimates of this quantity give
$\Omega_{\star}=(1.37 \pm 0.4)\times 10^{-3}$ from the 2dFGRS
\citep{Eke_et_al_05} and $\Omega_{\star}=(1.44 \pm 0.4)\times 10^{-3}$
from the SDSS \citep{Li_and_White_09}, assuming the Kennicutt and
Chabrier IMFs respectively. These estimates have a systematic
uncertainty of approximately a factor of 2 arising from the choice of
IMF.

\subsection{Cosmic downsizing}

Observations indicate that star formation in the most massive galaxies
observed today essentially concluded at $z \sim 1$
\citep[e.g.][]{Drory_et_al_03,Drory_et_al_05,Pozzetti_et_al_03_short,Fontana_et_al_04_short,Kodama_et_al_04},
and that their star formation rates were higher in the past than they
are at present
\citep[e.g.][]{Cowie_et_al_96,Juneau_et_al_05_short}. These findings
have been cited as a challenge to the $\Lambda$CDM cosmogony, since a
na\"ive view of galaxy formation within a hierarchical assembly
framework would have star formation shifting from smaller to larger
objects as the hierarchy builds up. This apparent conflict has spawned
terms such as ``downsizing'' and ``anti-hierarchical galaxy
formation''. Tentative evidence that these observations are compatible
with $\Lambda$CDM was first presented by \citet{Pearce_et_al_01},
using hydrodynamic simulations that featured star formation but no feedback mechanisms. \citet{Bower_et_al_06} and \citet{De_Lucia_et_al_06} convincingly demonstrated that the observations are entirely
compatible with their $\Lambda$CDM semi-analytic models. We now
explore whether our hydrodynamic simulations also exhibit downsizing.

\begin{figure}
  \includegraphics[width=\columnwidth]{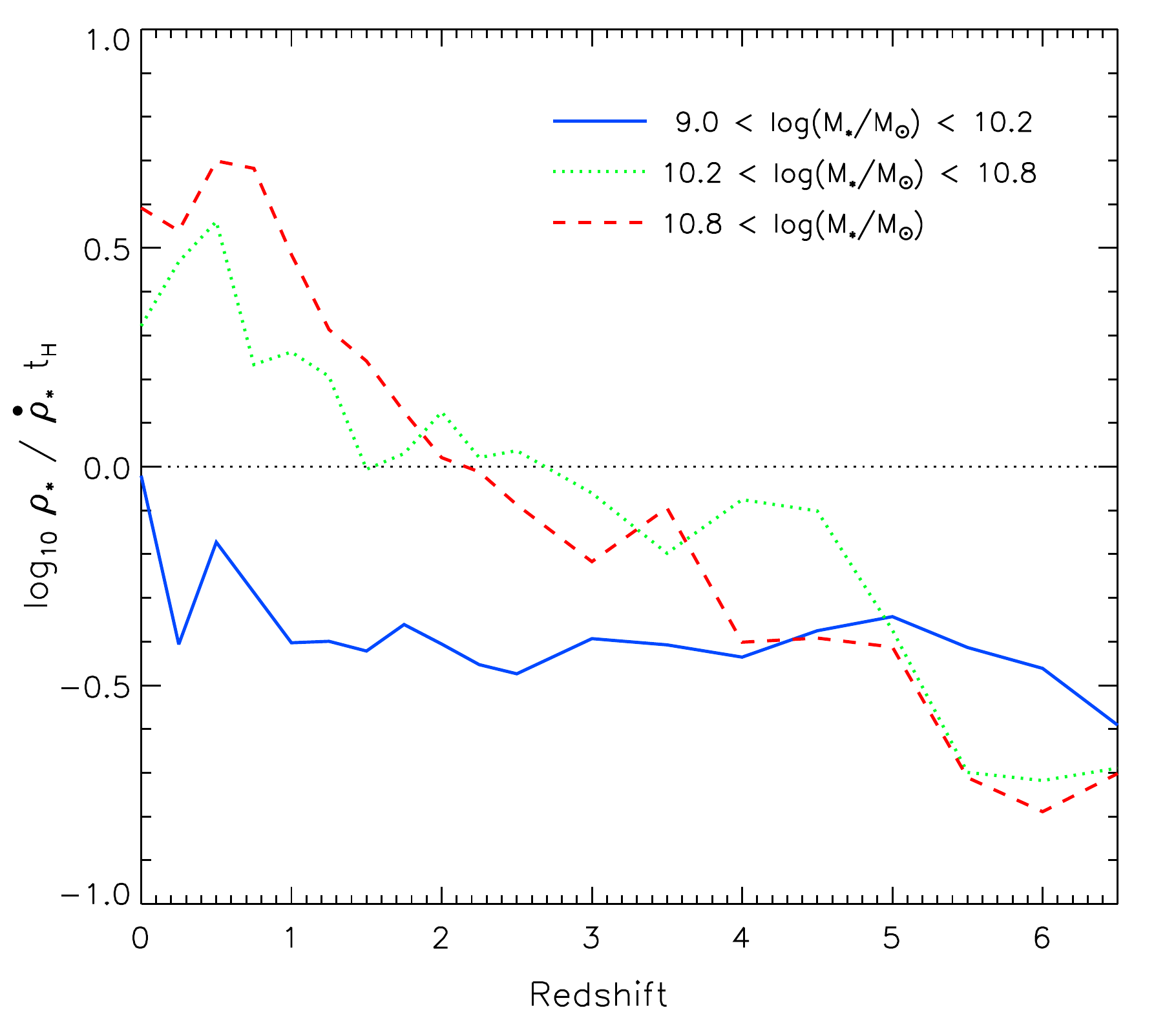}
  \caption{The ratio of past average star formation rate to the
current star formation rate at the redshift of observation, for three mass
    bins. The binning, matched to Fig.~7 of \citet{Bower_et_al_06},
    is done according to the stellar mass at the redshift of
    observation.  The dotted line at $\rho_\star / t_{\rm H} =
    \dot{\rho}_\star$ separates the regimes where i) on-going star
    formation dominates the current stellar mass (below), and ii)
    on-going star formation contributes a negligible fraction to the
    total stellar mass (above).}
    \label{fig:downsizing}
\end{figure}

One measure of downsizing, first highlighted by \citet[][their
Fig.~7]{Bower_et_al_06}, is the ratio of the past average SFR of
galaxies 
  of a given mass, $M_\star(z) / t_{\rm H}(z)$, where $t_{\rm H}$ is the Hubble time, to their SFR at the
  epoch of observation $\dot{M}_\star(M_\star,z)$. This is shown in a
  volume-averaged sense in Fig.~\ref{fig:downsizing}, for which
  galaxies have been drawn from all five intermediate-resolution 
  regions. Galaxies enter the regime above the horizontal dotted
  line that marks $\rho_\star / t_{\rm H} = \dot{\rho}_\star$ when they
  have assembled most of their stellar mass and their present star
  formation is adding only a minor contribution.

In common with results from the semi-analytic model of
\citet{Bower_et_al_06}, galaxies in \gimic\ exhibit a segregation by
stellar mass, such that the least massive galaxies continue to grow
until the present epoch, whilst more massive galaxies have essentially
concluded their star formation at $z>1-2$. Although this effect is
weaker in our simulations than in the semi-analytic model and in the
observational data, this is a significant result because our model
does not include AGN radio-mode feedback. Rather than causing the
downsizing effect directly, AGN feedback appears merely to exacerbate
it, pushing massive galaxies further into the regime where the past
average star formation rate dominates the current star formation
rate. AGN feedback does seem to be required, however, to prevent the
formation of the overly bright and blue galaxies that form in large
haloes in our simulations but which are not observed in nature. The
sort of downsizing discussed in this section results not from AGN
feedback but rather from the complex interplay of gas cooling, star
formation and feedback that develops as structure assembles
hierarchically.
 
\subsection{High-redshift star formation}

\begin{figure}
\includegraphics[width=\columnwidth]{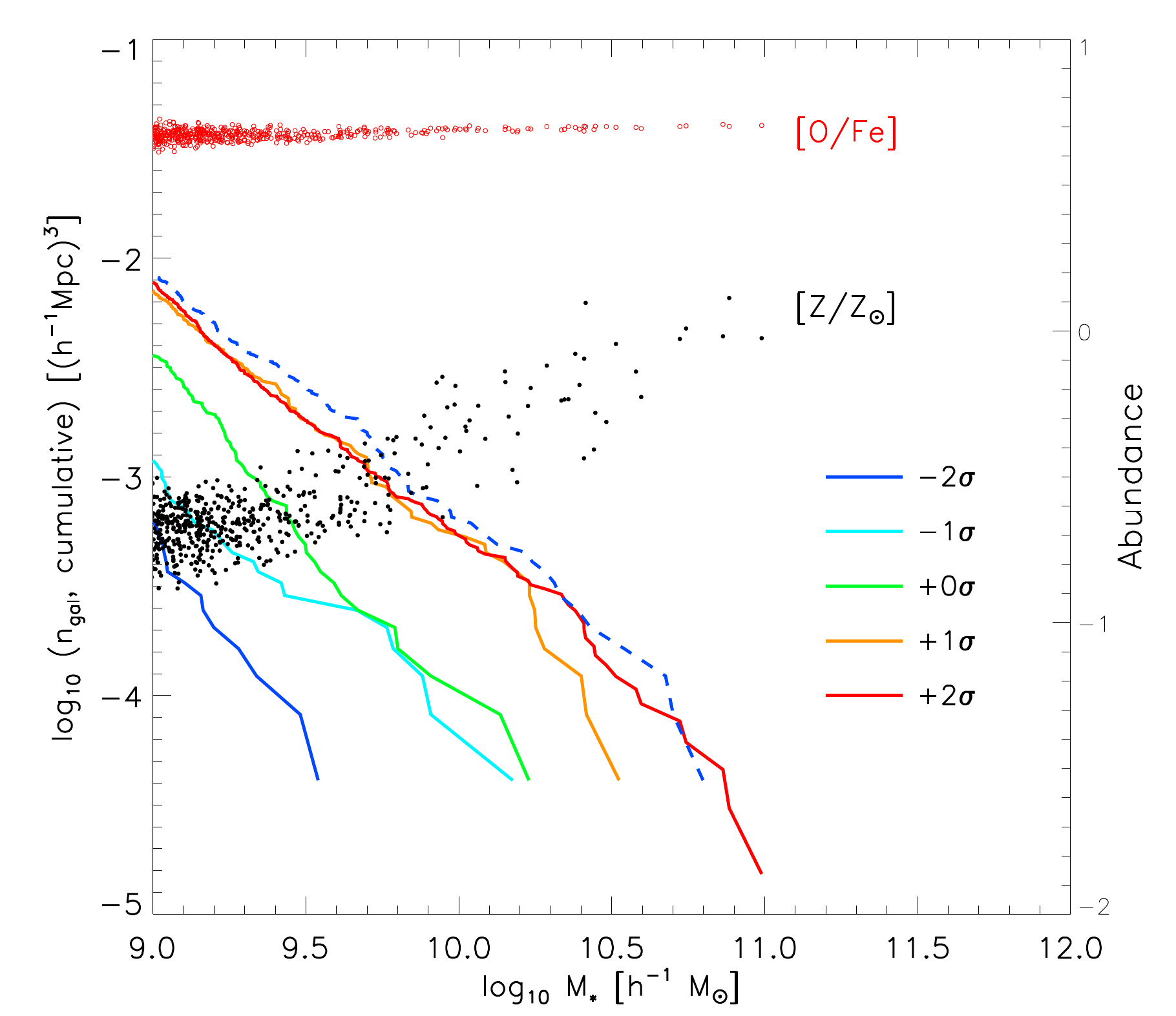}
 \caption{Cumulative number density of galaxies as a function of
 stellar mass at $z=6$ for the five \gimic\ regions (\textit{solid
 coloured lines}) at intermediate resolution. The number density for
 the \smm\ region is also shown at $z=3$ (\textit{dark blue dashed
 line}). The data plotted clearly illustrate the strong bias of massive
 stellar systems towards overdense regions. The most massive
 galaxies, drawn from the \spp\ region, have solar abundances
 (\textit{black points, right-hand y-axis}) and are highly
 overabundant in $\alpha$-process elements (\textit{red points,
 right-hand y-axis}).}  \label{fig:Highz}
\end{figure}

The observational data for $z\gtrsim 6$ are uncertain and differ by up
to a factor of ten between different studies; compare for example the
results of \citet{Bunker_et_al_06} and \citet{Bouwens_et_al_08}. Our 
high-resolution simulations favour a modest drop of $\sim 50$ per cent
in \sfrd\ from its peak at $z\sim 2$ to $z=6$, with most stars
forming in faint galaxies, which have star formation rates of
$\lesssim 0.1\Msunyr$. Such star formation rates are
typically an order of magnitude below the sensitivity limits of most
current surveys.  

The sky coverage of each $r=18\hMpc$
\gimic\ region is $320~{\rm arcmin}^2$ at $z=6$. Whilst this is comparable to
the sky coverage of the GOODS fields \citep{Dickinson_et_al_03}, it is
much greater than the $5-20~{\rm arcmin}^2$ of the Hubble Ultra Deep
Field \citep[HUDF,][]{Beckwith_et_al_06_short} in various passbands,
which is an important source for the detection and characterisation of
galaxies at $z > 6$. Since the mean thickness of a sphere is $4r/3$,
the comoving ``depth'' of each region is $\sim24\hMpc$. The expansion 
rate at $z=6$ is $H_0 = 930~h\HubUnits$, yielding a mean velocity
range of $\Delta v \sim 3190\kms$. This corresponds to $\Delta z \sim
0.07$ and therefore to a redshift ``window'' of $z=6\pm0.035$. Although
this is much narrower than the $\Delta z = 1.5$ estimated by
\citet{Bouwens_et_al_08} for their survey at $z=7$, the actual volume
probed with the pencil-beam geometry of the HUDF at these epochs is
similar to that of each \gimic\ region.  Our simulations indicate that
\sfrdz\ varies by up to an order of magnitude between regions of this
volume and so sample variance is a significant source of systematic
uncertainty in measurements of \sfrd\ at high redshift. This is
further compounded by the dominant contribution to \sfrd\ at these
epochs of low mass, inefficiently star-forming galaxies that are below
current detection limits.

The high-resolution simulations produce a slowly varying 
star formation rate density from $z=6$ to $z=9$. If the
escape fraction of ionising radiation from the small haloes that
dominate \sfrd\ at these epochs is high, say $\sim(25-80)$ per cent as
suggested by
\citet{Wise_and_Cen_09}, then these galaxies will be the dominant
contributors to the UV-background at high redshift, and thus may have 
been the main sources of the radiation that led to reionisation
\citep{Srbinovsky_and_Wyithe_08}. The faintness of each individual
source would reconcile the apparent dearth of detected sources of
UV-photons with the inferred ionisation state of the intergalactic
medium at $z\sim 6$ \citep{Bolton_and_Haehnelt_07}.

Although small galaxies dominate the star formation rate density, the
wide range of overdensities represented in the \gimic\ regions results
in massive stellar systems ($M_\star\sim 10^{11}\hMsun$) forming in
the simulations as early as $z=6$, as shown in
Fig.~\ref{fig:Highz}. The dark matter haloes that host these massive
galaxies are so strongly biased towards overdense regions that
galaxies of comparable mass do not form in the void region (\smm)
until $z\sim 3$ (\textit{dark blue dashed line}). These massive
galaxies are embedded in large, nearly spherical coronas of hot
($T\sim 10^7$ K) gas of comoving radius $\sim 0.3\hMpc$, but the
galaxies themselves are extremely compact (comoving radii of $\sim
3\hkpc$). The stars in these galaxies have near-solar metallicity and
are highly overabundant in the $\alpha$ elements produced in type II
SNe (such as oxygen), relative to those produced by type Ia SNe (such
as iron). They are relatively old (ages $\sim0.2\Gyr$) at this
redshift. Such galaxies are reminiscent of the $z\gtrsim 5$ massive
and evolved galaxy candidates found in the GOODS fields by
\citet{Wiklind_et_al_08}.

\section{The star formation properties of haloes}
\label{sec:halo_sfr}

We now turn to an exploration of the physical basis for the
environmental variation in the cosmic star formation rate density
described in Section~\ref{sec:sfrd}. We show that large-scale
environment has no impact upon the evolution of the star formation
rate in individual haloes. Instead, the strongly differing amplitudes
of \sfrd\ in the \gimic\ regions can be traced back to their different
mass functions.

We begin by considering the star formation rate in haloes of different
circular velocity and how this depends on the ``subgrid'' physics
assumed in the simulations. We compare the star formation properties
of the simulated haloes with analytic models for the regulation of
star formation by gas cooling, and show that winds effectively
regulate star formation up to a threshold value of the halo circular
velocity. In common with conclusions drawn in Section \ref{sec:sfrd},
from comparison with a semi-analytic model that includes AGN feedback,
we infer that efficient feedback in massive, hydrostatic haloes is
necessary to regulate the growth of the most massive galaxies and to
reconcile the cosmic star formation rate density at low redshift in
the simulations with observations.

\subsection{The specific star formation rate in haloes}
\label{sec:halo_ssfr}

\begin{figure*}
\includegraphics[width=0.8\textwidth]{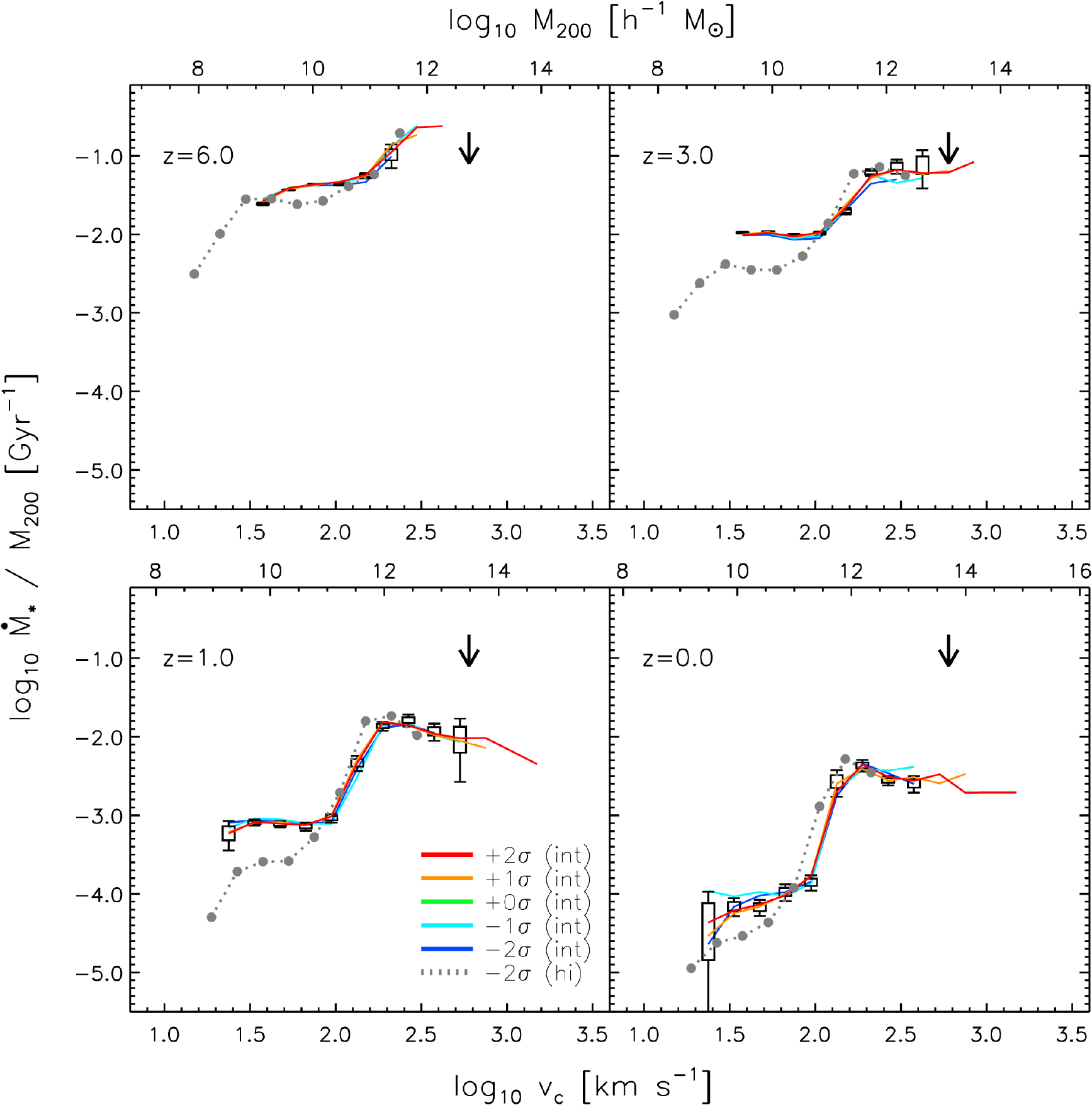}
  \caption{The ``halo specific star formation rate'', $\dot
    M_\star/M_{200}$, as a function of circular velocity ($v_{\rm c}$, bottom axis)
    or halo mass ($M_{200}$, top axis) for various redshifts. Different
    colours are for each intermediate-resolution \gimic\ region as
    indicated. Dotted grey lines are the results of the high-resolution
    $-2\sigma$ simulations. Box and whisker elements indicate one and
    two standard errors on the mean for the $0\sigma$ region. 
    The halo sSFR$-v_{\rm c}$ relations of the five \gimic\ regions are
    essentially indistinguishable and show a break at $v_{\rm c} \sim
    250\kms$, independently of redshift. The redshift
    evolution of the halo sSFR either side of this break reflects 
    the differing mechanisms acting to regulate star formation at
    different halo mass scales. The arrow at $600\kms$ indicates the
    launch speed of the winds.
}
    \label{fig:LargeScaleSSFR}
\end{figure*}

To characterise the star formation efficiency of haloes, we define the
\textit{halo} specific star formation rate, sSFR=$\dot
M_\star/M_{200}$, where $M_{200}$ is the total mass (baryons plus dark
matter) contained in a sphere of radius $r_{200}$. For clarity, we
shall retain the ``halo'' prefix to avoid confusion with the more common
definition of specific star formation rate that applies strictly to
\textit{galaxies}, $\dot M_\star/M_\star$.

In common with previous sections, we start by discussing the degree of
numerical convergence in the halo sSFR achieved by our
simulations. The mean halo sSFR as a function of circular velocity,
$v_{\rm c}=(GM_{200}/r_{200})^{1/2}$, and $M_{200}$ for haloes
identified in the five intermediate-resolution regions is shown at
various redshifts in Fig.~\ref{fig:LargeScaleSSFR}. The small box and
whisker elements denote one and two standard errors on the mean. Since
the relation for each region is virtually identical at each redshift,
convergence can be assessed by comparison with the relation from the
high-resolution realisation of the $-2\sigma$ region for all epochs
(\textit{dark blue solid} and \textit{grey dotted} curves).

For $v_{\rm c} \gtrsim 250\kms$, the halo sSFR in the $-2\sigma$
region is very similar for the two resolutions; for $100\kms \lesssim
v_{\rm c} \lesssim 250\kms$, the high-resolution simulation is only
slightly higher than the intermediate-resolution simulation. However,
below $100\kms$, the intermediate-resolution simulation clearly
overpredicts the halo sSFR (except at the highest redshift) and becomes
approximately independent of $v_{\rm c}$. Extrapolating from this
behaviour to haloes resolved with the same number of particles in the
high-resolution simulations, we conclude that these should give
reliable results for $v_{\rm c} \gtrsim 50\kms$. 

The halo sSFR-$v_{\rm c}$ relations for all \gimic\ regions are
virtually identical at every redshift. This similarity may be
unexpected, given the large variation in the \sfrd\ between the
regions; we defer to Section \ref{sec:multiplicity} a detailed
explanation of how these results are reconciled, in order to focus
initially on the origin of the halo sSFR-$v_{\rm c}$ relation. The
relation varies strongly with redshift, but drops rapidly below a
characteristic circular velocity that is essentially independent of
redshift. This drop occurs at $\sim 250\kms$, in the regime where the
high- and intermediate-resolution simulations agree well, and is
caused by the action of galactic winds. This scale is significantly
lower than the speed with which galactic winds are actually launched
($v_{\rm w}=600\kms$, indicated by an arrow) because the \textit{effective}
velocity of the winds is suppressed by drag forces imparted by the
dense gas of the ISM. \citet{Dalla_Vecchia_and_Schaye_08} demonstrated
that the winds become ineffective if the kick velocity falls below a
critical value that increases with the ISM pressure and hence with
halo circular velocity (or mass). Previous studies such as
\citet{Springel_and_Hernquist_03b} did not find this because the wind
particles were temporarily decoupled from the hydrodynamics and hence
could escape freely without experiencing any drag from the ISM. The
feature becomes more pronounced at later times when the halo sSFR
drops faster with time for lower masses.

Haloes of virial mass $M_{200}\sim 10^{9}\hMsun$ and circular velocity
$\sim 50\kms$ (that are well resolved in the $-2\sigma$
high-resolution simulation) have low star formation rates, $\dot
M_\star\lesssim 0.05\Msunyr$, at redshifts $z\sim 6$. Such objects are
ubiquitous at that time and they dominate the total star formation
rate. The halo sSFR drops rapidly at lower masses as haloes lose their
baryons due to a combination of stellar feedback and heating of their
gas by the UV-background. The
\textsc{Gadget2} simulations of \cite{Springel_and_Hernquist_03b}
yield quantitatively similar results, as do the AMR simulations of
idealised dwarfs described by \citet{Wise_and_Cen_09}. In the latter,
blow-out of baryons prevents continued star formation at masses below
$10^8\Msun$. In our simulations, the halo sSFR in these small haloes drops
rapidly by more than two orders of magnitude towards $z=0$.

Above a circular velocity of $\sim 250\kms$, where haloes are less
susceptible to galactic winds in our model, the halo sSFR is $\sim
10^{-1}\perGyr$, almost independently of mass above $z\sim 3$. The
level of this plateau drops by a factor $\sim 5$ to $z=1$. The
simulations of \citet{Springel_and_Hernquist_03b} yield a sSFR for a
halo of mass $10^{12}\hMsun$ of $10^{-1.7}\perGyr$ at $z=3$, a factor
of 3 below our value. They also show the halo sSFR at $z=1.5$, and the
drop from $z=3$ is a factor of 5, similar to the drop in our
simulations over the interval $z=1-3$. Whereas the halo sSFR is nearly
constant with mass for massive objects in our simulations, it
increases rapidly with halo mass in those of
\citet{Springel_and_Hernquist_03b}.

The halo sSFR plummets by a factor $\sim 50$ for a $10^{12}\hMsun$
halo between $z=6$ and $z=0$, and by an even greater factor 
at lower masses. The cause of this strong redshift evolution can be
understood from simple physical arguments.
\citet{White_and_Frenk_91} present an analytic model in which the halo
sSFR of a massive object of given circular velocity drops with
redshift as star formation becomes limited by the gas cooling time
rather than by its free-fall time.  Defining the cooling rate,
$\Lambda(u)$, via
\begin{equation}
\rho\dot{u} = -\Lambda n_{\rm H}^2, 
\end{equation}
where $n_{\rm H}= X\rho /m_{\rm H}$ is the hydrogen number density for
gas at density $\rho$, of which a mass fraction $X$ is hydrogen, and
$u$ is the thermal energy per unit mass, this cooling time is then
\begin{equation}
\tau_c(\rho) \equiv -{u\over \dot u} = -\frac{u}{\Lambda n_{\rm H}^2/\rho} \propto -\frac{u}{\Lambda\rho}.
\end{equation}
The cooling function depends on temperature, $T$. The virial temperature
of a halo is related to its circular velocity by:
\begin{equation}
T_{\rm vir} = {1\over 3}\,{\mu m_{\rm p}\over k_{\rm B}}\,v_{\rm c}^2 =
3.6\times 10^5\,{\rm K}\,\left({\mu\over 0.59}\right)\,\left({v_c\over 100~{\rm km}\,{\rm
    s}^{-1}}\right)^2,
\end{equation}
where $\mu$ is the mean molecular weight of the gas.

At high redshift, star formation is limited by the gas consumption
timescale since $\tau_{\rm c}$ becomes very small. So long as inverse
Compton cooling is unimportant, which is true at low
redshifts, the cooling time (and hence the halo specific star
formation rate) of haloes with virial temperature $\sim 10^7$~K
(corresponding to $v_{\rm c}=530\kms$), scales
$\propto\Lambda^{-1}\rho^{-1}$ which is approximately $\propto
\rho^{-1}=(1+z)^{-3}$ if $\Lambda$ does not evolve strongly. 

\begin{figure}
  \includegraphics[width=\columnwidth]{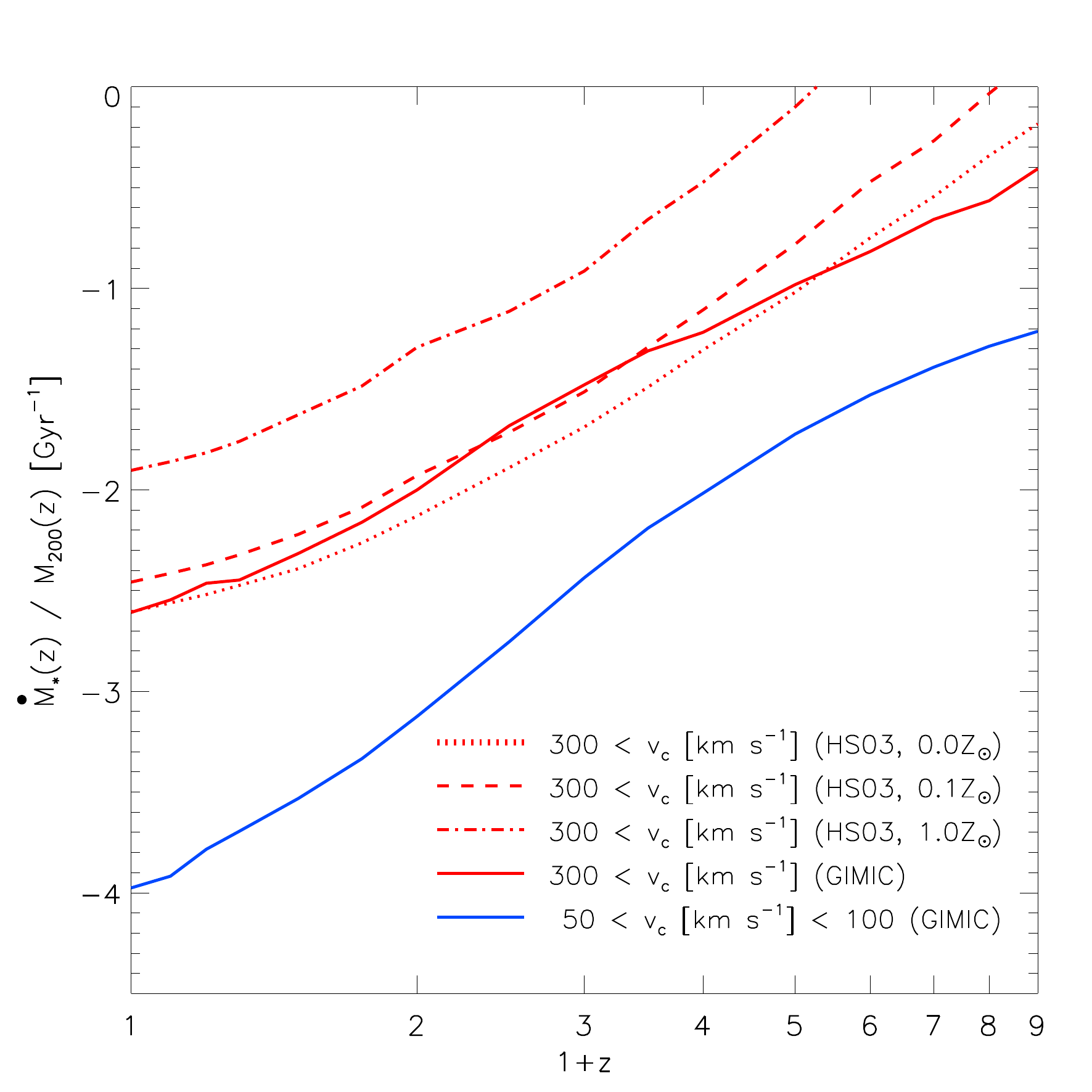}
  \caption{The mean halo specific star formation rate, $\dot
    M_\star/M_{\rm 200}$, as a function of redshift for haloes of
    circular velocity $v_{\rm c} < 300\kms$ (\textit{solid red line})
    and $v_{\rm c} = 50\kms$ (\textit{solid blue line}), drawn
    from the high-resolution \smm\ \gimic\ simulation. Results are
    compared with the analytic model of \citet[][labelled
    HS03]{Hernquist_and_Springel_03} for the massive haloes, assuming
    three different hot gas metallicities (primordial, \textit{dotted
      red}; $0.1\Zsun$, \textit{dashed red}; $1.0\Zsun$
    \textit{dot-dashed red}). The metallicity of the hot gas in the
    simulations is nearly constant at $\sim 0.1\Zsun$, and the model
    predicts the result for the massive halo in the simulation very
    well. At lower redshifts the drop in star formation follows that
    expected from the decrease in density. The analytic model is a
    poor description for low mass haloes (\textit{blue}), since it
    neglects the action of winds; the predicted halo sSFR is orders of
    magnitude higher than the simulated values, and so is not shown on
    the plot.}  \label{fig:sSFR}
\end{figure}

The evolution of the normalisation of the halo sSFR-$v_{\rm c}$
relation according to this model was explored analytically by
\citet{Hernquist_and_Springel_03}. Starting from pseudo-isothermal
initial gas profiles, their model yields specific star formation rates
for haloes of $T_{\rm vir}\simeq10^7\K$ that closely trace those in the
simulations of \citet{Springel_and_Hernquist_03b} at intermediate
redshifts ($7 \gtrsim z \gtrsim 2$). This model is compared to the high-resolution \gimic\ simulations in Fig.~\ref{fig:sSFR} where we have
computed the cooling rate taking into account the effect of an evolving
ionising background as given by \citet{Haardt_and_Madau_01} and
described by \citet{Wiersma_Schaye_and_Smith_09}. Note that this plot
does not represent the {\em evolution} of the halo sSFR of a given
object, but rather that of objects of a given circular velocity at {\em
different} redshifts.

The effect of metallicity on the sSFR is pronounced in the model, but
interestingly we find that there is, in fact, very little metallicity
evolution for the hot gas in the simulations, $Z\sim 0.1\Zsun$
from $z=9$ to the present. For this value of the metallicity, the model
works very well for massive haloes, indicating that the cooling time
controls star formation in these objects in the simulations, as
expected from the \citet{White_and_Frenk_91} calculation. (Note that
the inclusion of an efficient feedback mechanism in massive haloes,
such as energy injected by AGN, could alter the sSFR). The simulation
results and the \citet{Hernquist_and_Springel_03} model diverge at high
redshift because these authors chose not to include the gas consumption
timescale in this calculation; however, as \citet{White_and_Frenk_91}
show, taking account of this constraint would reduce the star formation
rate and improve the match at high redshift.

Efficient feedback depresses the halo sSFR in low-mass objects far
below the rate expected from the cooling time argument alone. The
accelerated drop-off (relative to the trend for massive haloes) at
$z\lesssim3$ is due to feedback from galactic winds which depletes the
baryon fraction in small haloes and increases the cooling time of the
remaining gas, quenching star formation. Winds therefore introduce a
second characteristic scale in galaxy formation in our model, that at
which star formation is quenched due to {\em the ejection of
baryons}. For our chosen wind speed of $v_{\rm w} = 600\kms$, this
scale is around $v_{\rm c} \sim 250\kms$ and depends on the details of
the wind implementation. This value is close to that advocated by
\citet{White_and_Frenk_91} as the scale separating massive haloes with
long cooling times from smaller haloes within which gas cools
efficiently.

\subsubsection{Comparison with semi-analytic models}

\begin{figure}
\includegraphics[width=\columnwidth]{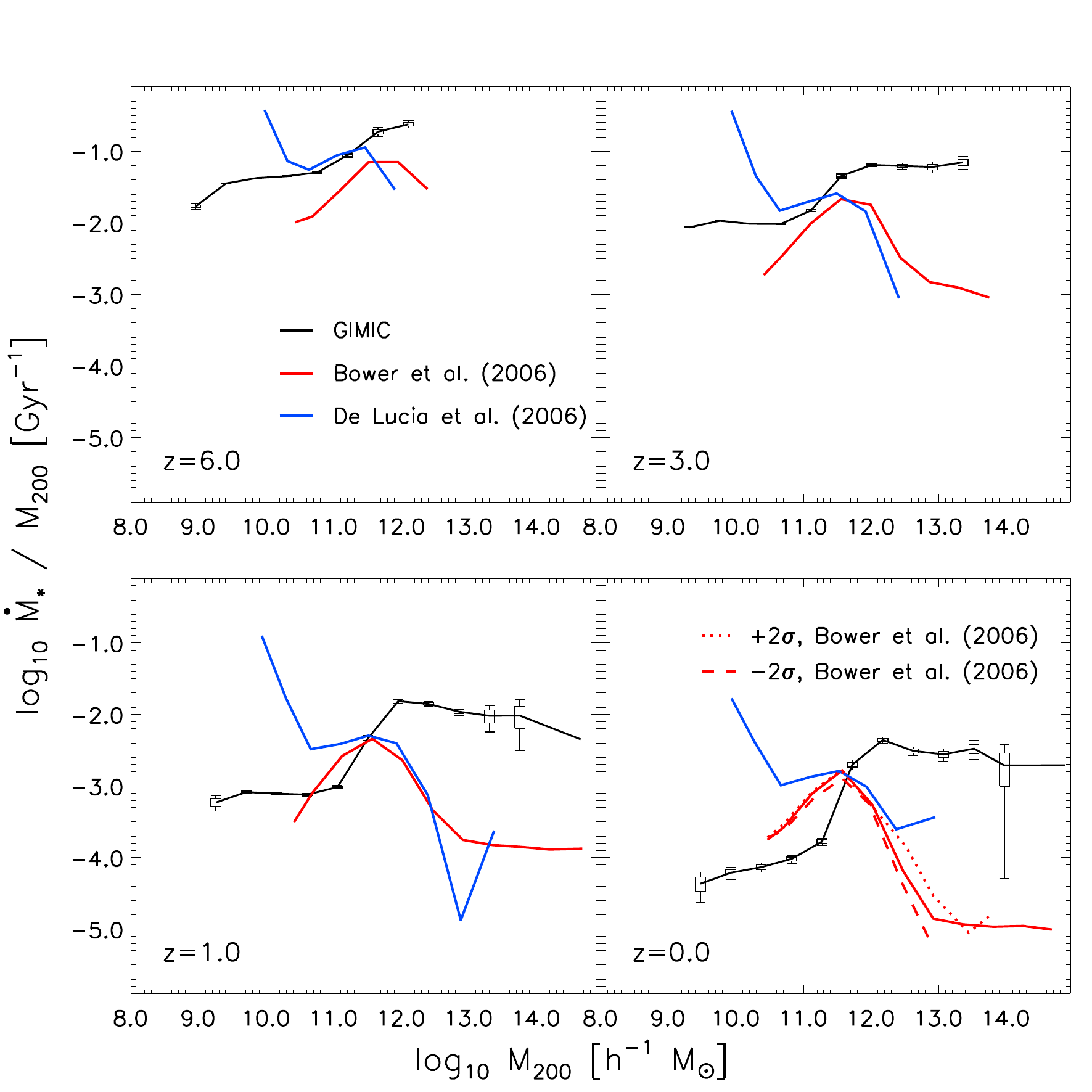}
\caption{The halo specific star formation rate, $\dot M_\star/M_{200}$,
  as a function of halo mass, $M_{200}$, (as in
Fig.~\ref{fig:LargeScaleSSFR}) for all \gimic\ regions combined,
compared 
  to the semi-analytic models of \citet[][\textit{red lines}]{Bower_et_al_06}
  and \citet[][\textit{blue lines}]{De_Lucia_et_al_06}. At $z=0$, results of the semi-analytic
  model are shown, limited to the \smm\ (\textit{red dotted curves})
  and \spp\ (\textit{dashed curves}) regions, rather than the entire
  Millennium Simulation volume; no difference is apparent, as in the
  hydrodynamic simulations.}  \label{fig:Bower}
\end{figure}

Although we defer a detailed comparison of the \gimic\ and
semi-analytic treatments of galaxy formation in the Millennium
Simulation to a later paper, it is instructive to carry out a limited
comparison here. We focus on the halo sSFR, $\dot M_\star/M_{200}$,
and compare, in Fig.~\ref{fig:Bower}, the results obtained by \citet{Bower_et_al_06} and
\citet{De_Lucia_et_al_06} with those from  \gimic.  
Note that the \gimic\ simulation extends to lower halo masses because
of its higher mass resolution.

For halo masses $M_{200}\gtrsim 10^{11}\hMsun$, the two
semi-analytic models give very similar results, particularly for
$z<3$. At smaller masses, however, there are substantial differences
between the semi-analytic models and also between these and the
simulations. These differences are largely due to different treatments
of feedback. The de Lucia et al. model has substantially higher sSFR
than the Bower et al. model reflecting the weaker feedback assumed in
the former; the feedback adopted in \gimic\ is intermediate between
the two semi-analytic prescriptions. For halo masses around $3\times
10^{11}\hMsun$, the \gimic\ results agree well with
the semi-analytic models. However, at higher masses \gimic\ produces
much higher star formation rates. The reason for this discrepancy is
simply that the semi-analytic models include feedback from the AGN radio-mode. As
explained in their respective papers \citep[see
also][]{Benson_et_al_03,Croton_et_al_06_short}, this ``radio mode feedback'' is required in
order to prevent the formation of overly massive galaxies in large
haloes. 

Finally, dotted and dashed lines in the bottom-right panel of Fig.~\ref{fig:Bower} compare the
\citet{Bower_et_al_06} halo sSFRs in the \spp\ and \smm\
\gimic\ regions. As in the \gimic\ simulations (see
Fig.~\ref{fig:LargeScaleSSFR}), the semi-analytic treatment finds
little, if any, dependence of the halo sSFR on the large-scale
environment.

\subsection{The multiplicity function of star formation and its
environmental dependence}
\label{sec:multiplicity}

We now return to an explanation of the physical reason for the strong
dependence of the cosmic star formation rate density, \sfrd, on
large-scale environment presented in Section~\ref{sec:large-scale
variation}. As shown in Fig.~\ref{fig:Madau_CONVERGENCE}, \sfrd\
varies by approximately an order of magnitude amongst the \gimic\
regions, which may seem surprising given the similarity of the
sSFR-$v_{\rm c}$ relation in each region.

The halo star formation rate density may be written in terms of the
specific star formation rate per halo as:
\begin{eqnarray}
\dot{\rho}_\star(z) &=& \int M_{200}
    \,{dN(M_{200},z)\over d\ln\,M_{200}}\,{\dot
     M_\star\over M_{200}}\,\,d\ln M_{200}\nonumber\\
                      &\equiv& \int g(M_{200},z)\,d\ln M_{200}\,,
\end{eqnarray}
where $dN(M_{200},z)/d\ln\,M_{200}$ is the mass function of dark
matter haloes. The function 
\begin{equation}
g(M_{200},z) = M_{200}\,{dN(M_{200},z)\over d\ln\,M_{200}}\,{\dot
    M_\star\over M_{200}}\,,
\label{eqn:g}
\end{equation}
is the {\em multiplicity function} of star formation, which describes
the star formation rate density in haloes of a given mass. As we
showed in Section~\ref{sec:haloes}, the halo mass function is a strong
function of environment, with the largest haloes forming only in the
densest regions. However, as we showed in Section~\ref{sec:halo_ssfr}
the star formation properties of a halo {\em at a given epoch},
including its specific star formation rate, $\dot M_\star/M_{200}$,
depend on its mass, not on its environment
(Fig.~\ref{fig:LargeScaleSSFR}), i.e. at any given epoch $\dot
M_\star/M_{200}$ is essentially independent of large-scale
environment. The dependence of \sfrd\ on large-scale environment
therefore reflects the dependence of the halo mass function on
large-scale environment. 

\begin{figure*}
  \includegraphics[width=\textwidth]{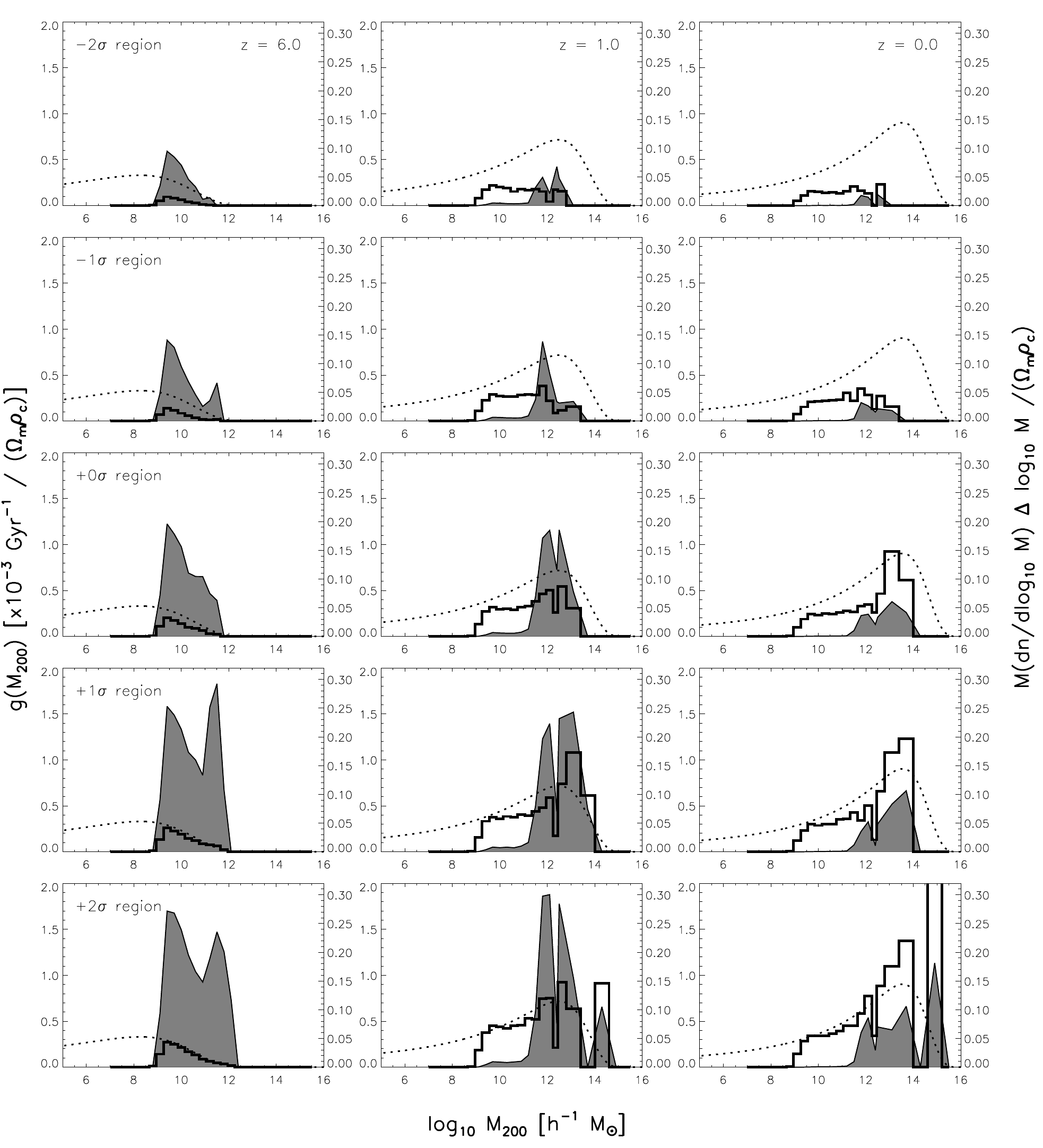}
  \caption{The multiplicity function, $g(M_{200},z)$, of star formation
    (shaded) as a function of region (rows) and redshift
    (columns). This function is the product of the halo sSFR and the
    halo multiplicity function, both of which are functions of redshift
    and halo mass. For reference, the halo multiplicity function of
    each region is reproduced from Fig.~\ref{fig:halo_mass_function}
    (\textit{black histogram, right-hand axis}), as is the multiplicity
    function derived from the universal mass function fit of
    \citet[][\textit{dotted curve, right-hand axis}]{Reed_et_al_07}.}
\label{fig:Multiplicity}
\end{figure*}

The multiplicity function of star formation, $g(M_{200},z)$, is shown
in Fig.~\ref{fig:Multiplicity} for each \gimic\ region at three
different redshifts; the star formation rate density of each region is proportional to the
shaded area. As highlighted by
\citet{Springel_and_Hernquist_03b}, the history of
$\dot{\rho}_\star(z)$ is shaped by the build-up of matter in haloes,
as described by $dN/d\ln M_{200}$, and the evolution of the specific
star formation rate, $\dot M_\star/M_{200}$. On the one hand,
advancing structure formation shifts an ever increasing fraction of
bound mass into more massive haloes, thereby pushing more and more gas
above the efficiency thresholds imposed by the photoionising
background and galactic winds. Conversely, the amplitude of $\dot M_\star/M_{200}$ is greatest at early times and drops steadily
over time. The combination of these effects yields a broad plateau in
$\dot{\rho}_\star$ at intermediate redshifts ($6\gtrsim z
\gtrsim 2$), from which it drops off towards lower and higher
redshifts.

Massive haloes are underrepresented in low density or ``void'' regions
and consequently star formation within voids is dominated by low-mass
haloes at high redshift (i.e. $z=6$), with the multiplicity function
of star formation peaking at $\sim 10^9\hMsun$. Only overdense regions have
sufficient numbers of massive haloes at these early times for massive
galaxies to contribute significantly to \sfrd; in these regions a
second peak develops at $\sim 10^{11}\hMsun$. By
$z=1$, massive haloes begin to be found even in voids and so the first
peak in $g$ moves to haloes of mass $\sim 10^{12}\hMsun$ in all regions with a second one developing around $\sim
10^{13}\hMsun$ in the high density regions. At the
present time, star formation is dominated by galactic haloes
($M_{200}\sim 10^{12}\hMsun$) in the void region, by
groups ($M_{200}\sim 10^{13-14}\hMsun$) in the
intermediate density regions, and by clusters ($M_{200}\sim
10^{15}\hMsun$) in the highest density region. It is
the absence of massive, efficient star-forming haloes in voids that so
severely inhibits their overall star formation rate density. This
explains the much greater difference in $\dot\rho_\star$ between the
underdense \smm\ and \sm\ regions, compared to that between the
overdense \sp\ and \spp\ ones.

Our simulations overestimate the specific star formation rates in
massive galaxies and haloes; hence the contribution of clusters is
likely to be exaggerated. Even so, the multiplicity function plot
illustrates the dynamic range problem faced by simulations of galaxy
formation: $\dot\rho_\star$ is dominated by $10^9\hMsun$ haloes at
$z=6$ but (in our simulations at least) by $10^{15}\,h^{-1}\,{\rm
M}_\odot$ haloes at $z=0$. As an example, the Millennium Simulation
does not resolve haloes below $2\times 10^{10}\hMsun$ and hence misses
the objects that (in our simulations) are the dominant contributors to
star formation at $z=6$. On the other hand, periodic simulations of,
for example, side length $L = 32\hMpc$ enclose a total mass of $\sim
10^{15}\hMsun$, and so cannot contain even a single cluster of the
type that dominates $\dot\rho_\star$ in the
\spp\ region. However, the fact that $\dot M_\star/M_{200}$ turns out
to depend mostly on halo mass and not on overdensity makes it
possible, in principle, to obtain accurate values for $\dot\rho_\star$
and its dependence on environment. This may be achieved by combining
accurate estimates of $\dot M_\star/M_{200}$ as a function of halo
mass with the dependence of the halo mass function on environment.

\subsection{Halo baryon fractions}

\begin{figure*}
\includegraphics[width=0.49\textwidth]{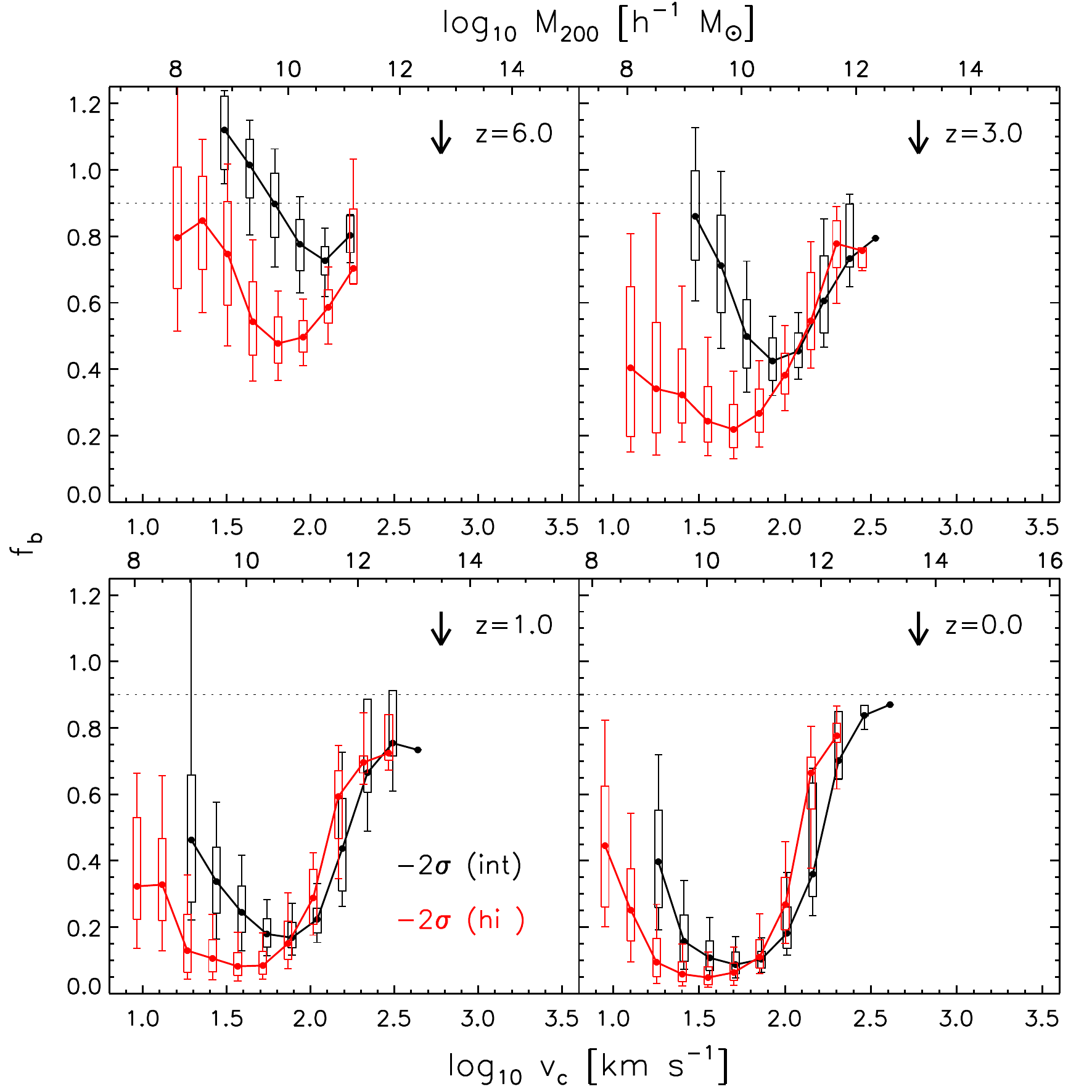}
\includegraphics[width=0.49\textwidth]{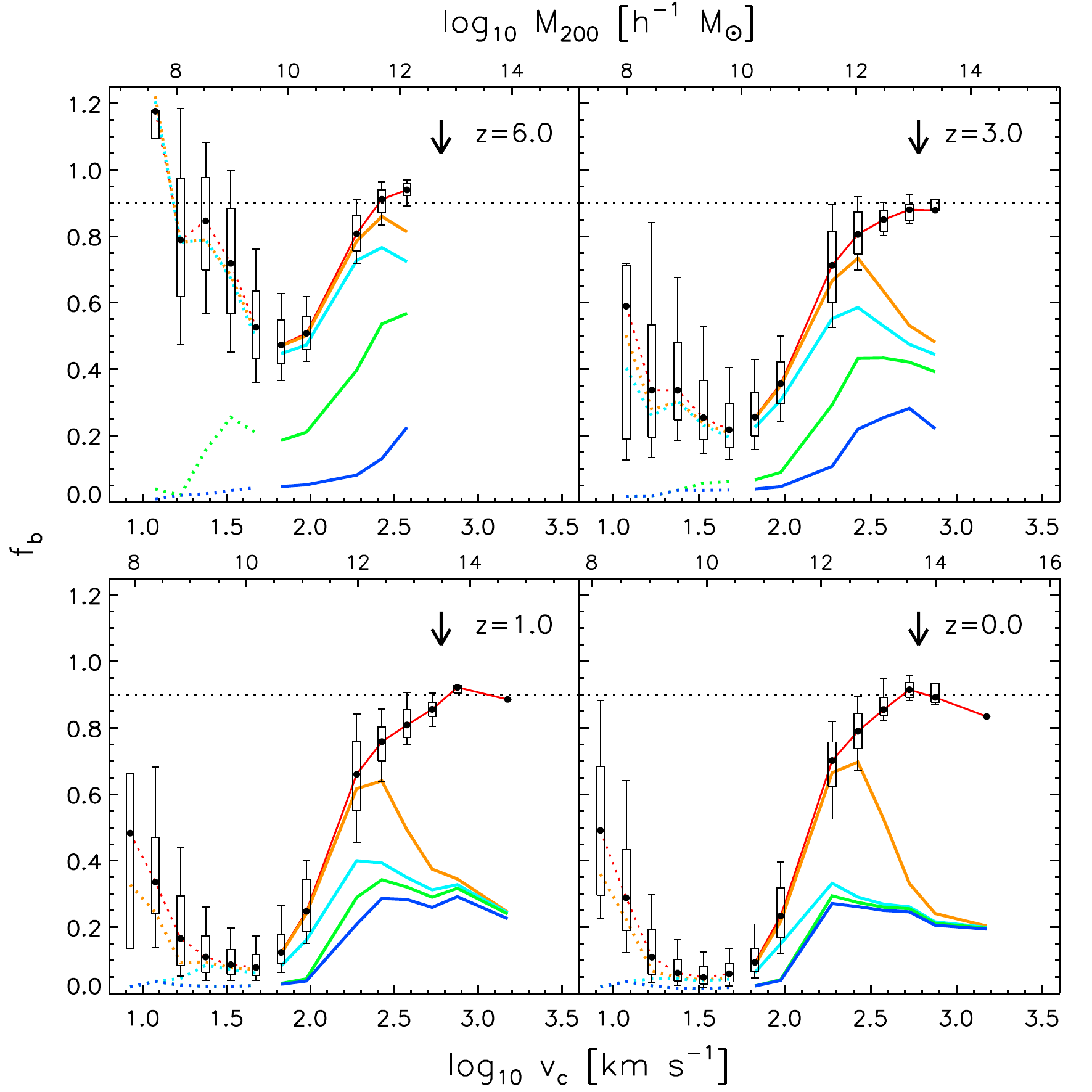}
\caption{Median baryon fraction, $f_{\rm b}\equiv (M_{\rm
    b}/M_{200})/(\Omega_{\rm b}/\Omega_{\rm m})$, of the simulated
  haloes. The dotted line indicates the 90 per cent level expected in
  the non-radiative regime; the arrows indicate the
  wind launch velocity. Box and whisker elements show the median and
  10$^{\rm th}$, 25$^{\rm th}$, 75$^{\rm th}$ and 90$^{\rm th}$
  percentiles of the total. \textit{Left:} resolution test, comparing
  the two realisations of the \smm\ region. Agreement between the
high- and intermediate-resolution runs for the total baryon fraction
is achieved at a circular velocity that 
  decreases with redshift; at $z=0$ this scale is $\sim
50\kms$. \textit{Right:} coloured curves decompose the baryon 
  fraction into the \textit{cumulative} contribution made by stars
  (\textit{dark blue}), equation of state gas (\textit{green}), cold
  gas (\textit{cyan}), warm gas (\textit{orange}) and hot gas
  (\textit{red}), in that order. Below $160\kms$, results are
  drawn from the high-resolution $-2\sigma$ region; above this scale
  they are drawn from all five intermediate-resolution regions. Dotted
  sections of the curves ($v_{\rm c} < 50\kms$) indicate
  the regime where we believe the high-resolution simulations to be
  unreliable.}
  \label{fig:BaryonFraction}
\end{figure*}

Stars comprise a relatively small fraction of the total baryonic
content of the Universe. In this subsection, we investigate how the
total baryonic content of haloes varies with halo circular velocity
and with time. This allows us not only to generate an inventory of the
baryons in our simulations but also to understand further the
processes responsible for the star formation properties discussed in
the previous subsection. As was the case for the sSFR, the baryon
fractions of haloes are essentially independent of environment and so,
when necessary, we combine haloes from all 5 \gimic\ regions to
improve the statistics. As before, we begin this discussion with a
convergence analysis.

The baryon fraction in units of the mean, $f_{\rm b}\equiv (M_{\rm
b}/M_{200})/(\Omega_{\rm b}/\Omega_{\rm m})$, is plotted as a function
of circular velocity (and halo mass, top axis) at several redshifts in
Fig.~\ref{fig:BaryonFraction}. To investigate convergence, we use the
intermediate- and high-resolution versions of the $-2\sigma$
region. The left-hand plot shows that approximate convergence is
achieved at a circular velocity of $v_{\rm c}\sim 100\kms$, the same
value at which, as shown in the preceding subsection, approximate
convergence in sSFR is also achieved. As before, extrapolating from
the behaviour seen in Fig.~\ref{fig:BaryonFraction}, we assume that
baryon fractions in the high-resolution simulation are reliable
down to $v_{\rm c}\sim 50\kms$ (that corresponds to the resolution
limit scaling with the number of particles in the intermediate- and
high-resolution simulations.) The rapid upturn in the baryon fraction
at $v_{\rm c} \lesssim 30\kms$ is thus likely to be an artefact of the limited
resolution.

The right-hand plot of Fig.~\ref{fig:BaryonFraction} shows how
the baryon content of a halo is made up from the following components: 
\begin{itemize}
\item stars
\item gas with $n_{\rm H} > 0.1\cmcubed$ (the value above which we impose an
 equation of state to mimic the multiphase ISM -- see \S2.2.)
\item cold gas 
\item warm gas
\item hot gas.
\end{itemize}

The cold and warm gas components are separated at $T=10^{5}\K$, whilst
the warm and hot gas components are separated at $T=10^{6.5}\K$. These
limits are chosen so as roughly to distinguish the warm-hot medium
from hotter X-ray emitting gas. The fractions of the different components are plotted
\textit{cumulatively} in the order listed above, so that the curve
showing the hot component is equivalent to the total baryon fraction.
For $v_{\rm c} < 160\kms$, we consider only the
(well resolved) haloes from the high-resolution $-2\sigma$ simulation;
for larger values of $v_{\rm c}$, we draw haloes from all five
intermediate-resolution runs to obtain a large sample of well resolved
haloes.

Massive haloes with circular velocity comparable to the launch speed of
winds or greater hold on to their baryons and, as shown in
Fig.~\ref{fig:BaryonFraction}, have $f_{\rm b}\sim 0.9$, as expected in the
non-radiative regime \citep[e.g.][]{Eke_et_al_98,Crain_et_al_07}. The
baryon fraction drops rapidly for haloes with $v_{\rm c} \lesssim
200\kms$.  These haloes have a virial temperature that
coincides with the peak in the cooling function. Efficient gas cooling
leads to high rates of star formation, but the associated galactic
winds are powerful enough to drive the gas out of the potential
wells. As a result these haloes have, in fact, {\em lower} specific
star formation rates (see Fig.~\ref{fig:LargeScaleSSFR}) and baryon
fractions than more massive haloes.

Whereas the total baryon fraction in haloes with $v_c\gtrsim 300\kms$
is nearly independent of redshift, the fraction in the star-forming
``equation of state'' gas phase drops dramatically from 35~per cent at
$z=6$ to a negligible fraction at $z=0$, when $\sim 75$ per cent is in
the form of hot gas. The overall baryon fraction of less massive
haloes, $v_{\rm c}\sim 100\kms$, drops significantly with
redshift.  Their stellar fractions increase with time and become
roughly constant for $z<3$, at around 5~per cent of the cosmic mean.
Because of their short gas cooling time, most of the gas in these low
mass haloes is initially in the cold phase. As star formation
proceeds, winds typically eject over half of the baryons that are
partially converted into warm and hot gas. Other mechanisms such as
ram-pressure stripping of gas and tidal stripping of stars can
contribute to the removal of baryons from these haloes.

\section{Summary and conclusions}
\label{sec:summary}

We have presented the first results of a programme of hydrodynamical
simulations of the formation of galaxies in a $\Lambda$CDM universe
being conducted by the Virgo consortium -- the
\textit{Galaxies-Intergalactic Medium Interaction Calculation} or
\gimic. The goal of the programme is to 
investigate the joint evolution of galaxies and the intergalactic
medium (IGM). The simulations follow the evolution of nearly spherical
regions of radius $\sim 20\hMpc$, drawn from the Millennium
Simulation \citep{Springel_et_al_05_short}. The regions have mean overdensities of
$(-2, -1, 0, +1, +2)\sigma$, where $\sigma$ is the root mean square
overdensity on the scale of the regions at $z=1.5$. The rest of the
Millennium Simulation volume is simulated at much lower resolution
using collisionless particles. This provides the correct tidal forces
on the high-resolution regions. 

The five \gimic\ regions were followed to $z=0$ at intermediate
resolution ($m_{\rm gas}=1.16 \times 10^7\hMsun$). Of the
high-resolution simulations ($m_{\rm gas}=1.45 \times 10^6\hMsun$),
the $-2\sigma$ region was continued to $z=0$, whilst the
$(-1,0,+1)\sigma$ regions were stopped at $z=2$. In the highest
resolution simulations the Jeans mass in the IGM after reionisation is
marginally resolved and thus these simulations can, in principle,
track the formation of all galaxies whose cooling is dominated by
H\textsc{i}.

Our simulation strategy has a number of advantages: (i)~it samples the
entire range of large-scale environments - including rare objects such
as massive cluster haloes and voids - allowing us to examine the
dependence of galaxy properties on large-scale overdensity; (ii)~it
allows accurate integration to $z=0$ since fluctuations on the scale
of the computational volume remain linear; (iii)~by suitably averaging
over the five regions, it allows an estimate of the statistical
properties of representative cosmological volumes.

The simulations were carred out using the \gadget\ code, a substantial
upgrade of \textsc{Gadget2} \citep{Springel_05} that includes:
\begin{enumerate}
\item a recipe for star formation designed to enforce a local
Kennicutt-Schmidt law \citep{Schaye_and_Dalla_Vecchia_08}; 
\item stellar evolution and the associated delayed release of 11
  chemical elements \citep{Wiersma_et_al_09};
\item the contribution of metals to the cooling of gas, computed
  element-by-element, in the presence of an imposed UV-background
\citep{Wiersma_Schaye_and_Smith_09}; 
\item galactic winds that pollute the IGM with metals and can quench star
  formation in low-mass haloes \citep{Dalla_Vecchia_and_Schaye_08}. 
\end{enumerate}
For these simulations we do not, however, follow the evolution of black holes or feedback
effects associated with them. 

The simulations produce the correct number density of galaxies with
mass greater than about $10^9 \Msun$. At $z=2$ the stellar mass
function is consistent with observations. At $z=0$ there are too many
galaxies of very low and very high mass and not enough of intermediate
mass. The excess at the massive end reflects the absence of a heating
mechanism, such as that often ascribed to AGN feedback, to prevent too
much gas cooling. At small and intermediate masses the discrepancy
with observations reflects inadequacies in our treatment of galactic
winds. For example, the paucity of galaxies with stellar mass
$\sim 10^{10}\Msun$ seems to be directly related to our simple
wind model.

In this paper we have concentrated on the star formation properties of
the simulations. These are governed by an interplay between accretion,
gas cooling, photoheating by the imposed ionising background and
galactic winds. The specific star formation rate in our simulations
is converged for $z\lesssim 6$ and we estimate that the value of
\sfrdz\ in the high-resolution simulations is reliable for $z < z_{\rm
reion} = 9$. Indeed, the reionisation process is ``visible'' in these
simulations as a depression of the star formation rate in low mass
haloes.  In all five \gimic\ regions, \sfrdz\ increases slowly with
time, reaches a broad maximum around $z\sim 2$, and then falls rapidly
by a factor of $\sim 6$ to the present day.  In general, the global
value of \sfrdz, obtained by a weighted average over the individual
regions, is in reasonable agreement with the observational compilation
of \citet{Hopkins_07}. However, the decline in \sfrd\ at low-redshift
in the simulation is not as pronounced as in the observations, largely
because of excessive star formation in massive haloes.

One of the main results of this paper is the strong environmental
dependence we find of the star formation rate density, \sfrdz, which
is manifest as a large variation amongst the five
\gimic\ regions. Although the shapes of the \sfrdz\ curves are
similar for all regions, the offset in amplitude between the
$-2\sigma$ to the $+2\sigma$ region is approximately an order of
magnitude.  The $0\sigma$ region closely tracks the global value
reconstructed from a weighted average of all five regions and is
typically a factor of 2 below the $+2\sigma$ region. Even when
normalised by total mass rather than volume, $\dot
\rho_{\star}$ still varies by more than a factor of 3 between the two
extreme regions.  This environmental dependence makes it difficult to
obtain reliable observational estimates, particularly at
high-redshift. For example, at $z=6$ the $18\hMpc$ \gimic\ spheres
correspond to 320 arcmin$^2$ on the sky, which is a considerably larger
area than the ($5-20$ arcmin$^2$) covered by the Hubble Ultra Deep
Field.

At a given redshift, we find that the specific star formation rate in
dark matter haloes, $\dot M_\star/M_{200}$ (the ``halo sSFR''), depends
only on halo mass, not on large-scale environment. At all redshifts, the specific
star formation is most efficient in haloes of circular velocity
$v_c=200-250\kms$. Above this value, the sSFR varies only little with
$v_c$, but below this value it drops rapidly due to the action of
galactic winds that deplete these haloes of baryons to below $\sim 10$
per cent of the cosmic mean by $z=0$. Above this critical value of
$v_c$, the sSFR is well described by the spherically symmetric cooling
model of \citet{White_and_Frenk_91}. Because of their generally higher
baryon fractions and gas densities, haloes of a given mass tend to
have higher sSFR at earlier times.

Since the halo sSFR is independent of environment, the reason for the
strong dependence of the star formation rate density with environment
is a strong dependence of the halo mass function on environment. There
are always more haloes and galaxies in the $+2\sigma$ region than in
the $-2\sigma$ region.  For example, the volume-normalised number density of galaxies with
stellar mass $> 10^{10}\hMsun$ at $z=0$ is a factor of $\sim3$ greater in the $+2\sigma$ than the $-2\sigma$ region.  In addition, the most massive
haloes are strongly biased towards regions of high overdensity. Since
the sSFR in our simulations is always dominated by high mass haloes
with $v_c=200-250\kms$, even when normalised to the total mass in each
region, the $+2\sigma$ region always has a higher \textit{specific}
star formation rate density than the $-2\sigma$ region. The combined
effect of halo number density and star formation efficiency results in
star formation at the present time that is dominated by galactic
haloes ($M_{200}\sim 10^{12}\hMsun$) in the void region, by groups
($M_{200}\sim 10^{13-14}\hMsun$) in the intermediate density regions,
and by clusters, $M_{200}\sim 10^{15}\hMsun$, in the highest density
region.

The plateau in the global $\dot{\rho}_\star(z)$ relation at $6\gtrsim
z \gtrsim 2$ is due to the balance between the advancing formation of
dark matter haloes and the decline in the sSFR in haloes of a given
mass (see Figs.~\ref{fig:LargeScaleSSFR} and
\ref{fig:Multiplicity}). Eventually, the latter dominates and \sfrd\
plummets in all regions. The lack of feedback effects associated with
the growth of central black holes may affect the quantitative, but
perhaps not the qualitative, behaviour of these trends.  This view is
supported by the agreement, to within a factor of $\sim 2$, between
the overall star formation rate densities inferred for the Millennium
Simulation from our simulations and from the semi-analytic models of
\citet{Bower_et_al_06,Croton_et_al_06_short} and
\citet{De_Lucia_et_al_06}, which do include the effects of AGN feedback.

In common with the conclusions of recent studies employing
semi-analytic modelling, we find that the growth of galaxies in our
simulations is segregated by stellar mass: low-mass galaxies continue
to grow at all epochs due to ongoing star-formation, whilst more
massive galaxies have largely concluded their star formation at
$z>1$. This apparent ``downsizing'' of the star formation activity
occurs in spite of the absence of AGN feedback in the simulations. 

The star formation rate density in the simulations is always dominated by
the progenitors of galaxies that today reside in the most massive
haloes (group/cluster haloes with $M\ge 5\times10^{12}\hMsun$). Yet,
these stars were predominantly formed either in dwarf haloes ($M\le
5\times10^{11}\hMsun$) before $z\sim 3.5$ or in galaxy-sized haloes
after that. Galactic winds are an effective mechanism for quenching
star formation to the observed levels in these smaller systems. The
integrated stellar mass density at $z=0$ agrees with observational
estimates to within a factor of $\lesssim 2$.

Some properties of the galaxies in the simulations can be compared
with observations. Galaxies form first in the high density
region. Between $z\simeq 9-6$, they are very compact (radii $\sim
3\,h^{-1}\,$ kpc), massive ($M_\star\sim 10^{11}\,h^{-1}\,{\rm
M}_\odot$), contain relatively old stars ($\sim 0.2$ Gyr old), have
high metallicity ($Z\sim {\rm Z}_\odot$) and are overabundant in
$\alpha$-elements. These galaxies are reminiscent of those recently
found in the GOODS fields by \citet{Wiklind_et_al_08}. 

In summary, the \gimic\ simulations provide a useful resource to
investigate the evolution of and interaction between galaxies and the
intergalactic gas in different large-scale environments. In this first
paper of a series, we have explored the star formation properties of
haloes and some of the properties of the galaxies that form within
them. In subsequent papers, we will examine further properties of both
galaxies and intergalactic gas.

\section*{Acknowledgements}   
\label{sec:acknowledgements}

The simulations were carried out using the HPCx facility at the
Edinburgh Parallel Computing Centre (EPCC) as part of the EC's DEISA
``Extreme Computing Initiative'', and with the Cosmology Machine at the
Institute for Computational Cosmology of Durham University. We extend
particular thanks to Lydia Heck and John Helly for expert HPC support,
and acknowledge fruitful conversations with members of the
\textsc{Virgo} and \textsc{Owls} consortia. RAC acknowledges support
from an STFC doctoral studentship. VRE is a Royal Society URF. CSF
acknowledges a Royal Society Wolfson Research Merit Award. PAT is
supported by STFC rolling grant ST/F002858/1. This work was supported
in part by an STFC rolling grant awarded to the ICC and by Marie Curie
Excellence Grant MEXT-CT-2004-014112.

\begin{appendix}
\medskip

\section{Methods in detail}

In this appendix we expand on the technical details of the generation
of the initial conditions for \gimic\ simulations. Secondly, we
discuss how we address the existence of an artificial boundary between
the high-resolution, baryon-filled region of each simulation, and the
external low-resolution region that is simulated with collissionless
dynamics. As part of this latter discussion, we describe our technique
for combining results from the five \gimic\ simulations to construct
estimates of properties of the Millennium Simulation as a whole.

\subsection{Generation of the initial conditions}
\label{sec:ics_appendix}

Here we provide a more detailed description of the generation of the
initial conditions, than the brief overview that was given in Section
\ref{sec:ics}. As described there, we aim to trace a representative sample of
the $(500~\hMpc)^3$ Millennium Simulation volume with five separate
simulations, and also to follow the evolution of rare structures such
as voids and clusters that are only found in such large periodic
volumes. Clearly, it is desirable to perform these simulations at
high resolution, and we aimed to have a sufficiently low particle mass
that the Jeans scale in the intergalactic medium (after the epoch of
reionisation) would be resolved in our ``high-resolution'' runs (which
have 8 times more particles than our 'intermediate-resolution'
runs). In practice, this limits the size of the regions to
$\sim18~\hMpc$. However, our additional constraint that the $+2\sigma$
region should be centred on a rich cluster at $z=0$ required the size
of this region to be increased to $25~\hMpc$; this is because the mass
of a sphere of mean density and radius $18\hMpc$ at $z=0$ is
$1.7\times10^{15}\hMsun$, a mass comparable to that of the largest
cluster in the Millennium Simulation. We describe the generation of
the initial conditions in two parts: i) the selection of the regions
to be resimulated, and ii) the generation of the particle distribution
and the displacement field.

\subsubsection{Selecting the five regions}

The four $(-2,-1,0,+1)\sigma$ spheres were selected from a target list
of $10^5$ randomly placed spheres of radius $18\hMpc$ in the
Millennium Simulation volume at $z=1.5$. Since by $z=1.5$ the
overdensity distribution has become significantly skewed
(i.e. non-Gaussian), we defined the overdensities of the
$(-2,-1,0,+1)\sigma$ regions as those that correspond in terms of the
rank ordering by overdensity, to a Gaussian distribution; the
overdensities for the $(-2,-1,0,+1)\sigma$ regions are
$(-0.407,-0.243,-0.032,0.241)$ respectively. From the complete list of
$10^5$ spheres a candidate list of centres was generated for each of
the regions by selecting all spheres with overdensities within 0.002
of the respective limits.

The procedure for the $+2\sigma$ sphere was different because of the
requirement that a rich cluster should form at its centre by
$z=0$. The starting point was to identify the particles that form the
126 most massive groups (that have a characteristic separation of
$\sim 100\hMpc$) in the Millennium Simulation volume at $z=0$,
identified by a friends-of-friends group finder \citep{DEFW85} with
linking length $b=0.2$. These particles were traced back to $z=1.5$
and their centres of mass used as centres for spheres of radius
$25\hMpc$.  The $+2\sigma$ sphere was then selected from a
randomly sorted list of twelve candidates that had an overdensity
close to the appropriate value.  Since the first cluster had a mass at
$z=0$ of $M_{200} = 4\times{10}^{14}\hMsun$, which is
rather low for a ``rich'' cluster, we chose the second sphere on the
list, with a mass of $M_{200}=8\times{10}^{14}\hMsun$ at
$z=0$. This cluster is relatively isolated, with no material in the
$25\hMpc$ belonging to another rich cluster.

Having selected the centre for the $+2\sigma$ sphere, and having made
candidate lists of centres for the four other spheres, we selected our
final five sphere centres from a list of all five centres generated as
follows. We selected a sphere, at random, from each of the candidate
lists of centres for the $(-2,-1,0,+1)\sigma$ regions and added the
chosen $+2\sigma$ sphere centre to form a quintet of candidate
centres. Each quintet was considered eligible for the final list if it
satisfied two criteria: (i) none of the five centres had one or more of
their $x$,$y$ or $z$ coordinates within $50\hMpc$ of the periodic
boundary of the simulation volume and (ii) each centre was at least
$200\hMpc$ away from the other four sphere centres in the
quintet. The first criterion was imposed to maintain the same
coordinate system as the Millennium Simulation, and also allow the use
of an intermediate-resolution particle-mesh (PM) force algorithm
implemented within our simulation code, which requires that the
isolated mesh used for this calculation should not cross the periodic
boundaries of the simulation volume. The second criterion was chosen to
prevent two or more spheres from being near neighbours, and to force the
five spheres to extend over a significant range of the Millennium
Simulation volume, so as to be relatively independent of each other.
It is possible to find many quintets satisfying these conditions; we
chose the first quintet generated by the code that used a pseudo-random
number generator to search for candidates. The centres and radii of all
five spheres are presented in Table \ref{tab:sim_params}.

In order to maximise the load and memory efficiency of the
intermediate-range PM algorithm in our code, which automatically
resizes to encompass all high-resolution particles, we chose to make
five separate sets of initial conditions, one for each of the regions
chosen; this is because the inclusion of all five spheres within a
single set of initial conditions would create an isolated mesh 
filling most of the simulation volume. Such a mesh would incur a
significant memory cost and offer little speed up relative to the main
PM algorithm. 

\subsubsection{Generating the particle load and applying the displacement field}

The procedure for making the initial conditions is identical for all
five regions and consists of two main stages: (i) the creation of a
multi-mass particle distribution to represent the uniform, unperturbed
particle distribution, and (ii) the recreation of the displacement
field used by the Millennium Simulation, with the addition of extra
short wavelength power in the resimulated regions, sampled from the
same power spectrum, down to the appropriate Nyquist frequency.  The
multimass uniform particle distribution is based upon a cubic mesh of
$320^3$ cells that encloses the entire simulation volume.  Each cell
of the mesh is treated independently. Particles, which at this point
are composite and represent both dark matter and gas, are placed in
each cell so that the mean density of all cells is equal to the matter
density of the Universe. A total of $n^3$ equal mass particles is
placed in each cell, arranged as a cubic grid with a spacing of $1/n$
of the cell size and with the centre of mass of all the particles
coinciding with the cell centre. The value of $n$ varied from 32, for
the high-resolution region encompassing the sphere itself, to unity
for cells that are distant from the sphere and are needed only to
provide the correct tidal forces on the region of interest.

The value of $n$, as a function of cell position, was determined by the
following algorithm: all of the particles from the Millennium
Simulation inside or within $0.5\hMpc$ ($1\hMpc$ for the
$+2\sigma$ sphere) of a particular \textsc{Gimic} sphere at $z=1.5$
were traced back to the grid cells from which they originated at very
high redshift. Any cell occupied by one or more of these particles is
labelled as a high-resolution cell.  A check is made to determine if
the high-resolution cells form a single contiguous region, where
contiguous cells are defined as those sharing a face. If there were
non-contiguous cells present, then any cells contiguous to a
high-resolution cell are additionally defined as high-resolution cells,
and this process is continued until all high-resolution cells are
contiguous. A set of boundary layers is then defined as follows: level
1 boundary cells are defined as those cells adjacent (i.e. sharing a
face, an edge or a vertex) to a high-resolution cell.  Level 2 boundary
cells are then adjacent to level 1 boundary cells, and so on until
level 9, which in addition includes all cells that have not yet been
assigned. 

We then realise the zoomed initial conditions at two resolutions; since
the original Millennium Simulation represents our low-resolution basis,
we refer to the two \textsc{Gimic} realisations as high- and
intermediate-resolution. For the high-resolution realisation, we used
$n=32$ in the high-resolution cells, while the neighbouring boundary
cells had $n = (20, 15, 10, 8, 6, 4, 3, 2, 1)$ for the boundary levels
1-9 respectively. This yields a mass for the composite (i.e. dark
matter and gas) particles in the high-resolution region of
$8.08\times10^6\hMsun$. For the intermediate-resolution
realisation, we effectively halve the value of $n$ in each level (odd
values are rounded up), yielding initial conditions with composite
particle masses a factor of eight greater than the high-resolution
case, $6.46\times10^7\hMsun$. We therefore resolve haloes of
mass $10^{12}\hMsun$ (comparable to the Milky Way's halo)
with $\sim19,000$ particles at intermediate-resolution, and
$\sim150,000$ at high-resolution. The total number of particles used in
each simulation is presented in Table \ref{tab:sim_params}.

Having created the particle distribution, a displacement field was calculated
for each particle using the techniques for making resimulation
initial conditions described in \citet{Power_et_al_03} and
\citet{Navarro_et_al_04_short}. The displacement consists of two parts;
on large scales the same mode amplitudes and phases were used as in
the Millennium Simulation, whilst short wavelength power down to the
Nyquist frequency was added in a cubic region of dimension
$50-76\hMpc$ containing all of the high-resolution cells.  The
break between the long wavelength and short wavelength power was
chosen such that only wavenumbers $\vec{k}=(k_{x},k_{y},k_{z})$
satisfying $\rm{max}$$(|k_{x}|,|k_{y}|,|k_{z}|) < \pi h~$Mpc$^{-1}$
were retained from the Millennium Simulation. This scale is sufficient
to ensure that haloes resolved with at least a few hundred particles
in the Millennium Simulation are present in the \textsc{Gimic}
resimulation. A large mesh of $2048^3$ cells was used for all Fourier
transforms to minimise interpolation errors.

Finally, the high-resolution composite particles were split into gas
and dark matter particles, with masses appropriate to yield a cosmic
baryon fraction of $\Omega_{\rm b}/\Omega_{\rm m} = 0.045/0.25 =
18~$per cent. The gas and dark matter particles were offset in each
dimension from their common centre of mass by a mass-weighted fraction
of one interparticle separation.

\subsection{Sampling \& combining regions}
\label{sec:sampling}

The use of zoomed initial conditions complicates the interpretation of
simulations because of the presence of an artificial boundary around
the high-resolution region. Inside it, baryons are modelled explicitly
with SPH particles; outside it, dark matter and baryons are
represented by composite particles whose dynamics are
collisionless. Clearly, the collisionless region should be excluded
from any analysis, but the existence of the boundary has other
indirect effects. For example, the absence of gas beyond the boundary
artificially reduces the mass of gas that might accrete onto
peripheral haloes. In addition, the lack of external pressure allows
the easy escape of galactic outflows into the vacuum of the boundary
region.

In the analyses presented in this paper that measure properties on a
particle-by-particle basis, we consider only those particles that are
within $18\hMpc$ (comoving; $25\hMpc$ for the $+2\sigma$ region) of
the centre of mass of all baryonic and high-resolution dark matter
particles; in the case of analyses where we consider the properties of
haloes or galaxies, we consider those whose centre-of-mass is within
this volume. For $z\gtrsim 1.5$ this automatically leaves a small
boundary layer of SPH and high-resolution dark matter particles
external to the sphere for ``safety'', since a padding region was
included in the initial conditions by design. However, at lower
redshifts the morphology and comoving volume of each region deviates
slightly from that of the sphere that defined them at $z=1.5$. This is
due to the varying overdensity of each region, since the underdense
regions expand and the overdense regions contract in comoving
space. However, we have explicitly checked that haloes towards the
edge of the volume do not differ significantly from those at the
centre, for instance in terms of their baryon fraction, star formation
rate or metallicity, and so we do not consider this an issue for the
analyses we present. In addition, the use of a simple and
time-invariant sampling geometry (a sphere of constant comoving
radius) allows various quantities, such as the star formation rate
density $\dot{\rho}_\star$, to be normalised by volume in an
unambiguous fashion.

\begin{figure}
  \begin{center}
\includegraphics[width=\columnwidth]{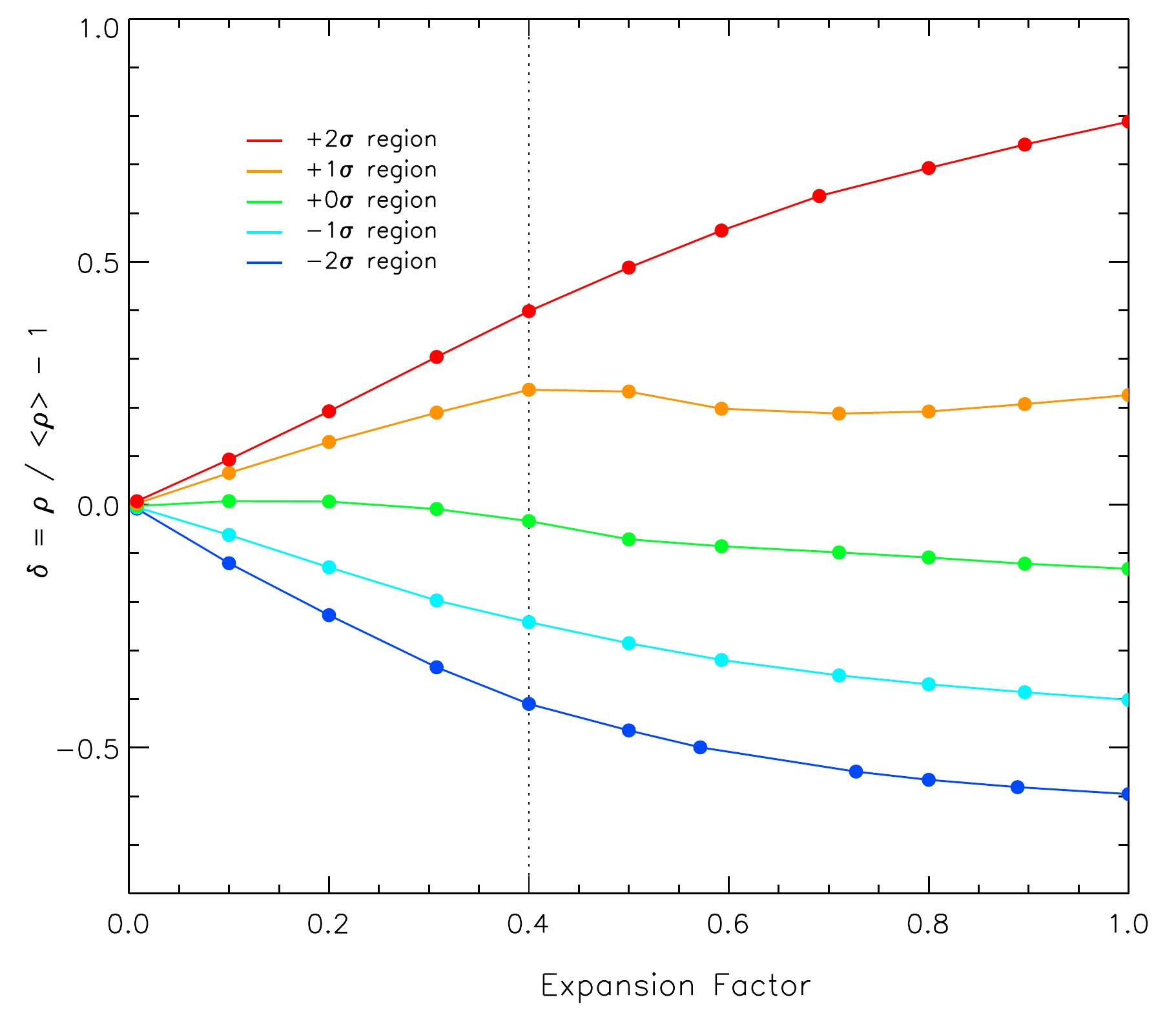}
    \caption[The evolution of overdensity in the five spherical regions
      at intermediate resolution]{The overdensity evolution of the five spherical regions
      at intermediate resolution. The vertical dotted line denotes the
      epoch at which the regions were selected ($z=1.5$). Notice the drop, after $z=1.5$, in
      overdensity of the $+1\sigma$ region; this results from a
      massive halo with high peculiar velocity leaving the spherical
      region.}
    \label{fig:overdensity_evolution}
    \end{center}
\end{figure}
In Fig.~\ref{fig:overdensity_evolution} we present the overdensity
evolution of the spherical regions within each simulation. At early
times, the regions necessarily have the mean density of the Universe,
but their enclosed overdensities clearly diverge as they evolve. This
figure highlights the advantage of zoomed initial conditions as a means
to probe varying environments, since a periodic volume maintains
$\delta=0$ at all epochs. The dip in the overdensity of the $+1\sigma$
region at $a\simeq0.5~(z\simeq1)$ results from the departure of a
high-mass halo with a large peculiar velocity from the spherical
``analysis'' region. 

At $z=1.5$ the distribution of overdensities over spherical regions of
size comparable to the GIMIC regions is close to Gaussian. Ideally, to
obtain the universal average of a quantity which is a function of
overdensity, it is necessary to integrate over overdensity with a
Gaussian weighting. Because the \gimic\ simulation provides only 5
overdensities. $(-2,-1,0,+1,+2)\sigma$. we approximate the integral as
a sum and choose the weights for the five regions so that the
summation correctly integrates any odd polynomial of overdensity and
even polynomials up to and including a quartic power in overdensity.
These weights are given in the second column of
Table~\ref{tab:weighting}.  An additional complication is that the
$+2\sigma$ region has a different size, so an additional weight is
needed to compensate for the larger volume (Table~\ref{tab:weighting},
third column). The actual weights are given in the fourth column
Table~\ref{tab:weighting}. 

\begin{table}
  \begin{center}
    {\small
      \begin{tabular}{|l|l|l|l|}
	\hline\hline
	Sphere                 & Summation   & Volume     & Net        \\
	                       & weight      & weight     & factor     \\
	\hline						
	$-2\sigma$             & 1/12       & 1           & 1/12       \\
	$-1\sigma$             & 1/6        & 1           & 1/6        \\
	$\phantom{+} 0\sigma$  & 1/2        & 1           & 1/2        \\
	$+1\sigma$             & 1/6        & 1           & 1/6        \\
	$+ 2\sigma$            & 1/12       & $(18/25)^3$ & 486/15625  \\
	\hline
      \end{tabular}
    }
    \caption{Numerical weights applied to each \gimic\ region
      when extrapolating statistics to the whole $(500\hMpc)^3$
      Millennium Simulation volume. The net factor (final column) is
      the product of the summation weight assigned to each region
      (first column) and the volume weighting (second column). The
      latter is applied to account for the larger volume of the
      $+2\sigma$ region.}
    \label{tab:weighting}
  \end{center}
\end{table}

\begin{figure}
\includegraphics[width=0.48\textwidth]{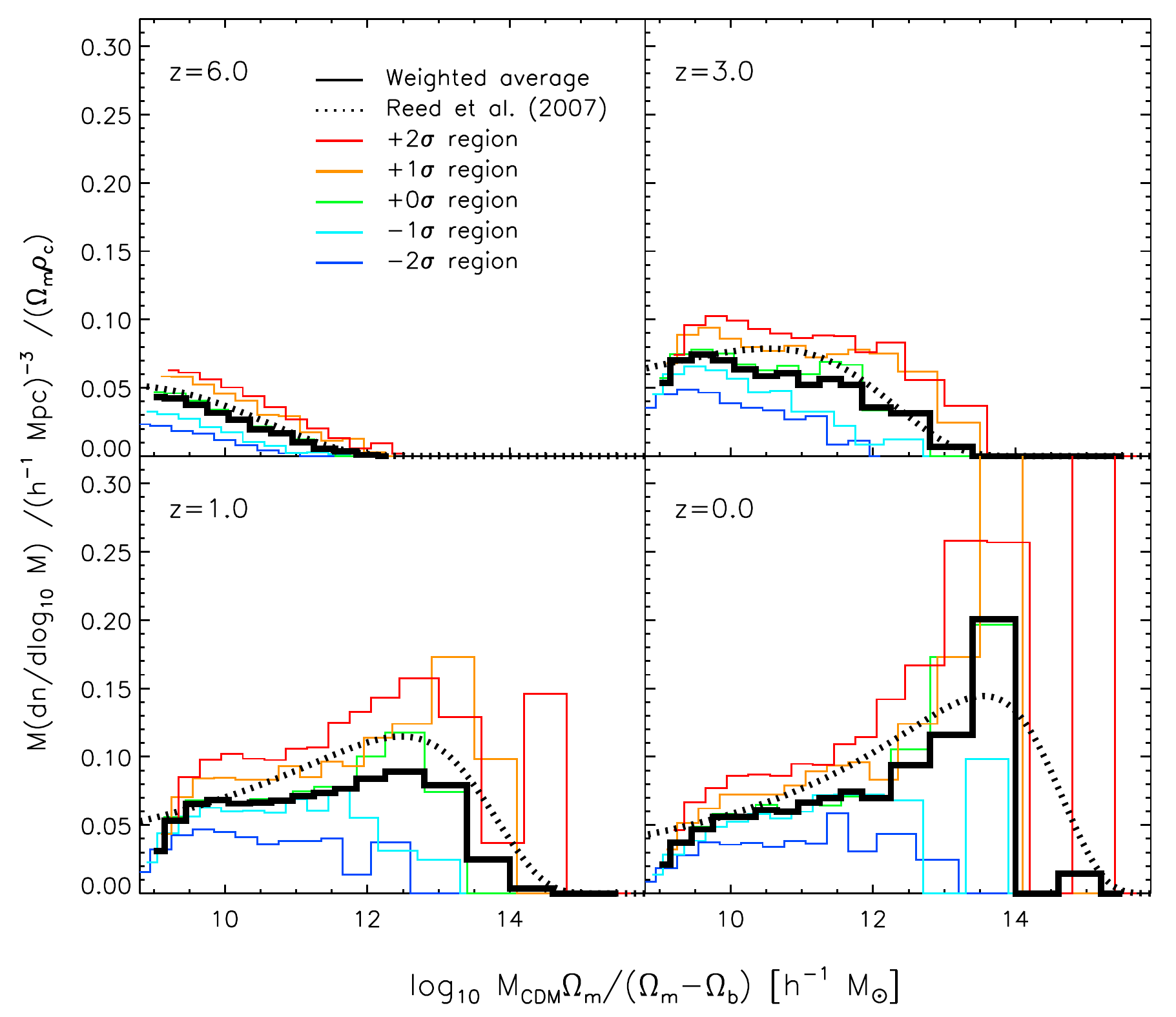}
  \caption{The halo multiplicity function in the five
    regions at intermediate resolution. The black histogram is the halo
    multiplicity function yielded by our weighting procedure used to
    generate estimates for the Millennium Simulation volume from the
    \gimic\ regions. For comparison, we show the multiplicity function
    derived from the \citet{Reed_et_al_07} mass function
(\textit{dotted line}).}
    \label{fig:halo_multiplicity_function}
\end{figure}

These weights can only be approximate because the overdensity
distributions are not exactly Gaussian, and strictly the regions only
correspond to $(-2,-1,0,+1,+2)\sigma$ deviations from the mean at
$z=1.5$.  In Fig.~\ref{fig:halo_multiplicity_function}, we show the
\textit{halo multiplicity function}, defined as the fraction of all
mass bound into haloes of mass in the interval $(\log M,\log M+d\log
M)$. This description of the halo population is attractive since it
has a smaller dynamic range than the halo mass function, and so more
clearly highlights the \textit{differences} between the halo
populations of the
\gimic\ regions. Inspection of this plot shows that our weighted average
(\textit{black histogram}) constructed from the individual regions
(\textit{coloured histograms}) recovers the halo multiplicity function (and
therefore the halo mass function also) of the Millennium Simulation
(\textit{dotted line}) reasonably well at all times. It is this fact that
ultimately justifies our weighting scheme. 

\end{appendix}

\label{lastpage}

\end{document}